\documentclass[fleqn,usenatbib]{mnras}

\usepackage{newtxtext,newtxmath}

\usepackage[T1]{fontenc}
\usepackage{caption}
\usepackage{booktabs}
\usepackage{longtable}
\usepackage{subcaption}

\DeclareRobustCommand{\VAN}[3]{#2}
\let\VANthebibliography\thebibliography
\def\thebibliography{\DeclareRobustCommand{\VAN}[3]{##3}\VANthebibliography}

\usepackage{graphicx}	
\usepackage{amsmath}	

\usepackage{xcolor}
\usepackage{ulem}

\newcommand{\rev}[1]{#1}

\title[The resolved radio source counts to 5$\,\mu$Jy]{Deblending the MIGHTEE-COSMOS survey with XID+: The resolved radio source counts to $S_{1.4}\approx 5\mu$Jy}

\author[Malefahlo ~et~al.]{Eliab Malefahlo\thanks{eliabmalefahlo3@gmail.com}$^{1,2}$, Matt J.~Jarvis$^{3,2}$, Mario G. Santos $^{2,4}$, Catherine Cress$^{1}$, Daniel J.B. Smith$^{5}$,
\newauthor
Catherine Hale$^{3,6}$, Jos\'e Afonso$^{7,8}$, Imogen H. Whittam$^{3,2}$, Mattia Vaccari$^{9,10,11}$, Ian Heywood$^{12, 3, 13}$, 
\newauthor
Shuowen Jin$^{14, 15}$, Fangxia An$^{16,9}$
 \\
\\
$^{1}$UNISA Centre for Astrophysics and Space Sciences, College of Science, Engineering and Technology, University of South Africa, Cnr Christian de Wet Rd \\ \quad and Pioneer Avenue, Florida Park, 1709 Roodepoort, South Africa\\
$^{2}$Department of Physics \& Astronomy, University of the Western
Cape, Private Bag X17, Bellville, Cape Town, 7535, South Africa\\
$^{3}$ Astrophysics, Department of Physics, University of Oxford, Keble Road, Oxford, OX1 3RH, UK\\
$^{4}$ South African Radio Astronomy Observatory (SARAO), 2 Fir Street, Observatory, Cape Town, 7925, South Africa\\
$^{5}$ Centre for Astrophysics Research, University of Hertfordshire, Hatfield, Herts, AL10 9AB, UK\\
$^{6}$ Institute for Astronomy, University of Edinburgh, Royal Observatory Edinburgh, Blackford Hill, Edinburgh, EH9 3HJ, UK \\
$^{7}$ Instituto de Astrofísica e Ciências do Espaço, Universidade de Lisboa, OAL, Tapada da Ajuda, PT1349-018 Lisbon, Portugal\\
$^{8}$ Departamento de Física, Faculdade de Ciências, Universidade de Lisboa, Edifício C8, Campo Grande, PT1749-016 Lisbon \\
$^{9}$ Inter-University Institute for Data Intensive Astronomy, Department of Astronomy, University of Cape Town, 7701 Rondebosch, Cape Town, South Africa \\
$^{10}$ Department of Astronomy, University of Cape Town, Rondebosch, Cape Town, 7701, South Africa \\
$^{11}$ INAF - Istituto di Radioastronomia, via Gobetti 101, 40129 Bologna, Italy \\
$^{12}$ SKA Organization, Jodrell Bank, Lower Whitington, Macclesfield, SK11 9FT, UK \\
$^{13}$ Centre for Radio Astronomy Techniques and Technologies, Department of Physics and Electronics, Rhodes University, PO Box 94, Makhanda, 6140, South Africa \\
$^{14}$ Cosmic Dawn Center (DAWN), Denmark \\
$^{15}$DTU Space, Technical University of Denmark, Elektrovej 327, 2800 Kgs. Lyngby, Denmark\\
$^{16}$ Yunnan Observatories, Chinese Academy of Sciences, Kunming 650216, People’s Republic of China}
\date{Accepted XXX. Received YYY; in original form ZZZ}

\pubyear{\the\year{}}

\begin{document}
\label{firstpage}
\pagerange{\pageref{firstpage}--\pageref{lastpage}}
\maketitle

\begin{abstract}
Deep radio continuum surveys provide fundamental constraints on galaxy evolution, but source confusion limits sensitivity to the faintest sources. We present a complete framework for producing high-fidelity deblended radio catalogues from the confused MIGHTEE maps using the probabilistic deblending framework XID+ and prior positions from deep multi-wavelength data in the COSMOS field. To assess performance, we construct MIGHTEE-like simulations based on the Tiered Radio Extragalactic Continuum Simulation (T-RECS) radio source population, ensuring a realistic distribution of star-forming galaxies and active galactic nuclei (AGN) for validation. Through these simulations, we show that prior catalogue purity is the dominant factor controlling deblending accuracy: a high-purity prior, containing only sources with a high likelihood of radio detection, recovers accurate flux densities and reproduces input source counts down to $\sim 3\sigma$ (where $\sigma = $ thermal noise). On the other hand, a complete prior overestimates the source counts due to spurious detections. Our optimal strategy combines the high-purity prior with a mask that removes sources detected above $50~\mu$Jy. Applied to the $\sim$1.3\,deg$^2$ area of the MIGHTEE-COSMOS field defined by overlapping multi-wavelength data, this procedure yields a deblended catalogue of 89,562 sources. The derived 1.4\,GHz source counts agree with independent P(D) analyses and indicate that we resolve the radio background to $\sim 4.8\,\mu$Jy. We also define a recommended high-fidelity sample of 20,757 sources, based on detection significance, flux density, and goodness-of-fit, which provides reliable flux densities for individual sources in the confusion-limited regime.
\end{abstract}


\section{Introduction}

The bright radio sky, at Giga-Hertz frequencies and flux densities above a few millijanskys (mJy), is dominated by a powerful population of radio-loud Active Galactic Nuclei (AGN; e.g., \citealt{Kellermann1989,Best2012, Hardcastle2020,Whittam2022}). In these sources, the observed emission is predominantly non-thermal synchrotron radiation, produced as relativistic electrons, accelerated within powerful jets launched by central supermassive black holes (e.g., \citealt{BlandfordRees1974,Blandford1979}). The statistical distribution of extragalactic sources is an important tool in cosmology, quantified by the differential source counts ($dN/dS$), which measure the number of sources per unit solid angle as a function of flux density.  At these bright, mJy-to-Jy flux densities, where radio-loud AGN dominate, the normalized counts are relatively flat due to the rarity of these luminous AGN, whose sky density is very low (e.g., \citealt{Condon1984}).  However, below a few mJy, they exhibit a well-established upturn, a clear signature that the source population is evolving and was significantly more numerous (e.g., \citealt{Condon1984,Windhorst1985}). This upturn marks a shift in the nature of the dominant radio source population.

The emerging population that dominates the faint sub-mJy and $\mu$Jy radio sky is now understood to be composed of two main classes of object: star-forming galaxies (SFGs) and radio-faint AGN (e.g., \citealt{Jarvis2004,Wilman2008,deZotti2010,White2015,White2017,Padovani2016, Smolcic2017}). In SFGs, the radio emission at Giga-Hertz frequencies originates from processes linked to massive stars: it is dominated by non-thermal synchrotron radiation from cosmic-ray electrons accelerated in supernova remnants (the endpoints of massive stellar evolution), and a smaller contribution from thermal free-free emission from HII regions, which are clouds of interstellar gas ionized by the intense ultraviolet radiation from young, massive stars \citep{Condon1992, Bell2003, Murphy2011, Kennicutt2012}. Because these processes are a direct consequence of recent massive star formation, the radio luminosity serves as a well-established tracer of the star formation rate (SFR, \citealt{Condon1992}). Furthermore, because radio waves are unaffected by the dust that obscures optical and ultraviolet light, these observations provide one of the least biased measures of the cosmic star formation history, an important quantity for galaxy evolution studies \citep{Karim2011, Jarvis2015, Novak2017,Ocran2020, Matthews2021b,Malefahlo2022, Wang2024b,Matthews2024}. 

The origin of radio emission in the radio-quiet AGN population, is a subject of ongoing debate. Defined as AGN that lack the powerful, large-scale relativistic jets characteristic of their radio-loud counterparts, their radio emission is orders of magnitude weaker. One theory argues that the emission, while weak, is still intrinsically linked to AGN activity, such as small-scale jets, coronal emission from the accretion disk, or outflows and winds \citep{White2015, White2017, Hwang2018, Rankine2021,Wang2025}. An alternative theory is that this emission is not directly related to the AGN but is instead dominated by star formation in the host galaxy \citep{Kimball2011, Condon2013, Bonzini2013, Gurkan2015, Malefahlo2020}
. The current consensus suggests that both processes likely contribute, but their relative importance as a function of key parameters like black hole accretion rate, stellar mass, and redshift remains a key open question in galaxy evolution \citep[e.g.][]{Bonfield2011,Hickox2014,Panessa2019,Macfarlane2021,Yue2024,Kondapally2025}.

The radio emission originating from star-formation processes is empirically found to correlate tightly with the infrared luminosity of galaxies, a relationship first discovered in the far-infrared (FIR; \citep{Helou1985,Yun2001} and now commonly referred to as the total infrared--radio correlation (IRRC; e.g., \citealt{Bell2003}). This near-linear correlation holds over several orders of magnitude in luminosity and is known to extend to high redshifts, although its precise evolution with cosmic time remains an active area of research (e.g., \citealt{Jarvis2010, Magnelli2015, Read2018, Delhaize2017,Delvecchio2021}). Given the strong correlation between far-infrared emission and the star-formation rate (SFR), this leads to the radio-SFR relation \citep[e.g.][]{Yun2001,Bell2003,Gurkan2018,Smith2021} and the potential to study star-formation in the highest redshift galaxies \citep[e.g.][]{Whittam2025}.
There is a growing body of evidence that demonstrates that it is not only pure star-forming galaxies that follow this relation; the host galaxies of most low-luminosity and radio-quiet AGN are also observed to lie on or very near the IRRC (e.g.,
\citealt{Moloko2025}).

The current generation of radio telescopes has now reached the sensitivity, resolution and survey speed needed to probe the $\mu$Jy sky in detail. Key instruments include the MeerKAT telescope in South Africa \citep{Jonas2009,Booth2009}, the Australian SKA Pathfinder (ASKAP; \citealt{Johnston2008}), and the Low-Frequency Array (LOFAR; \citealt{vanHaarlem2013}). These are enhanced by upgraded arrays like the Giant Metrewave Radio Telescope (uGMRT \citealt{Gupta2017}) 
, with the future Square Kilometre Array Observatory (SKAO; \citealt{Braun2015}) and the Next Generation Very Large Array (ngVLA; \citealt{Murphy2018}) set to push sensitivity even further. With their ability to combine sub-$\mu$Jy sensitivity with arcsecond resolution over wide fields, these facilities are designed specifically to address key questions in galaxy evolution: to build a complete census of the cosmic star formation history, disentangle the faint SFG and radio-faint AGN populations, and investigate the faint-end slope of the radio source counts.

Despite these instrumental advances, deep surveys run into the instrumental confusion: 
the surface density of faint sources becomes so high that individual objects overlap/blend 
due to the telescope's beam, making direct detection and photometry unreliable \citep{Condon1974, Zwart2014}. 
There are two strategies that have been developed to push beyond confusion.
The first is the probability of deflection or P(D) analysis, which 
can recover the underlying source counts but does not provide individual source flux densities \citep{Scheuer1957, Condon1974, Condon2012, Vernstrom2014, Matthews2021}. The second
approach is prior-driven deblending, which uses high-resolution ancillary data to model and partition the confused maps, predominantly for submillimetre and far-infrared data \citep{Tibshirani1996, Zou2006, Magnelli2010, Bethermin2010, Roseboom2010, Chapin2011, Lee2013, Wang2014, Hurley2017, Jin2018, Wang2024}. The prior-driven approach has multiple methods, from image-plane model fitting like `super-deblending' \citep{Liu2018,Jin2018} to probabilistic Bayesian frameworks like XID+ \citep{Hurley2017}, which samples the full posterior probability distribution of flux densities. However, the success of these probabilistic frameworks is highly sensitive to the construction of the prior lists. 

In this paper, we develop and validate a complete framework based on this probabilistic approach to produce a deblended radio source catalogue from the confused survey maps of the MeerKAT International GHz Tiered Extragalactic Exploration (MIGHTEE) survey \citep{Jarvis2016,Heywood2022}. MIGHTEE is a large survey project using the MeerKAT telescope to map deep extragalactic fields, with the first data release \citep[DR1;][]{Hale2025} providing the data for this work. We use realistic simulations to optimize our methodology before applying this validated strategy to the MIGHTEE observations of the Cosmic Evolution Survey (COSMOS; \citealt{Scoville2007}) field. In Section \ref{sec:data}, we describe the MIGHTEE radio data and the ancillary multi-wavelength catalogues used to construct our prior lists. Section \ref{sec:xidplus} provides a brief overview of the XID+ deblending framework. In Section \ref{sec:sim_validation}, we present the analysis of realistic simulations used to validate and optimize our deblending strategy. In Section \ref{sec:application}, we apply this optimized strategy to the MIGHTEE-COSMOS field, presenting the final deblended flux catalogue and radio source counts. Finally, we summarize our key findings and conclusions in Section \ref{sec:conclusions}.
Throughout this paper, we assume a flat $\Lambda$CDM cosmology with $H_0 = 70\,\mathrm{km\,s^{-1}\,Mpc^{-1}}$, $\Omega_\mathrm{M} = 0.3$, and $\Omega_\Lambda = 0.7$. Furthermore, we define the radio spectral index ($\alpha$) by the power-law relation between flux density ($S_\nu$) and frequency ($\nu$), given by $S_\nu \propto \nu^\alpha$.

\section{Data}
\label{sec:data}

This section details the primary radio data, the ancillary multi-wavelength catalogues, the star-galaxy classification process, and the final definition of the master galaxy sample used for subsequent analysis.

\subsection{MIGHTEE DR1 Radio Data}
We use radio data from the first MIGHTEE data release, DR1 \citep{Hale2025}. This release includes deep imaging over 20 deg$^2$ of the COSMOS, XMM-LSS, and CDFS fields. For this work, we focus specifically on the MIGHTEE-COSMOS field, which covers a total area of 4.2 deg$^2$. The DR1 data for this field were produced from 22 individual pointings, totaling 139.6 hours of on-target time at L-band ($\sim$1.2--1.3 GHz). The final mosaicked images are released at two resolutions: a lower resolution ($\sim$8.9 arcsec) prioritizing sensitivity and a higher resolution ($\sim$5.2 arcsec). The thermal noise is $\sim$1.6 $\mu$Jy/beam (lower resolution) and $\sim$2.2 $\mu$Jy/beam (higher resolution), with a median measured total noise across the mosaic of $\sim$3.5 $\mu$Jy/beam \citep{Hale2025}. We adopt two distinct noise symbols throughout: $\sigma$ for the thermal noise and $\sigma_{\rm Tot}$ for the total noise, which includes confusion noise $\sigma_{\rm conf}$. The depth and area of these maps make them rich with faint sources, leading to source confusion that motivates this work. We apply the deblending technique to the deeper low-resolution data, which suffers more from confusion \citep{Heywood2022}.

\subsection{Ancillary Multi-wavelength Data}
\label{sec:ancillary_data}

The probabilistic deblending performed in this work requires an input prior list of potential source positions. We construct our master source list using the deep multi-wavelength catalogue presented by \citet{Varadaraj2023}, which builds upon the data compilation of \citet{Adams2021}. This catalogue is built upon the UltraVISTA DR4 near-infrared (NIR) data \citep{McCracken2012}, providing deep \textit{Y, J, H}, and \textit{K$_s$} band photometry. This is complemented by optical coverage from the Hyper Suprime-Cam (HSC) Subaru Strategic Program PDR3 \citep{Aihara2022}, as well as data from the Canada-France-Hawaii Telescope (CFHT; \citealt{Gwyn2012}) and \textit{Spitzer}/IRAC 3.6 and 4.5\,$\mu$m photometry \citep{Fazio2004, Mauduit2012}.

To obtain the necessary physical properties for our prior construction, we use the COSMOS2020 photometric catalogue \citep{Weaver2022}. This is a comprehensive, panchromatic photometric catalogue covering the COSMOS field, updated from the COSMOS2015 release \citep{Laigle2016}. It includes new, deep imaging data from several key surveys: U-band data from the CLAUDS survey \citep{Sawicki2019}, \textit{g, r, i, z, y}-band data from the HSC Subaru Strategic Program PDR2 \citep{Aihara2019}, \textit{Y, J, H, $K_s$}-band data from the UltraVISTA DR4 survey \citep{McCracken2012}, and updated \textit{Spitzer}/IRAC data from the Cosmic Dawn Survey \citep{EuclidCollaboration2022}. Source detection was performed on a deep, $\chi^2$ \textit{izYJHKs} detection image. We use the \texttt{FARMER} catalogue, which was generated using the Tractor software, where the catalogue is limited to the UltraVISTA footprint \citep{Weaver2022}. \rev{The \texttt{FARMER} catalogue is preferred over the \texttt{CLASSIC} catalogue because it uses profile-fitting photometry via The \texttt{Tractor} software, which provides superior deblending in crowded regions compared to the aperture-based photometry used in \texttt{CLASSIC} \citep{Weaver2022}.}

From this catalogue, we use the photometric redshifts and inferred total infrared luminosities ($L_{\mathrm{IR}}$ = rest-frame 8--1000\,$\mu$m luminosity) derived using the \texttt{EAZY} code \citep{Brammer2008}. This $L_{\mathrm{IR}}$ value is an inferred quantity derived from SED fitting based on the principle of energy balance. Specifically, this calculation is based on the method implemented in the EAZY code, which uses Flexible Stellar Population Synthesis (\texttt{FSPS}) templates \citep{Conroy2009} to model the starlight and estimate the energy absorbed by dust. This absorbed energy is then assumed to be re-radiated in the FIR, following the spectral templates of \cite{Magdis2012}. \rev{We note that the COSMOS2020 catalogue relies solely on optical-to-mid-infrared photometry (up to IRAC 4.5\,$\mu$m) and does not include far-infrared or longer wavelength data in the SED fitting process \citep{Weaver2022}.}

The catalogue includes a \texttt{FLAG\_COMBINED} column, which identifies sources within a clean area of  1.278\,deg$^2$ with full HSC, VISTA, and \textit{Spitzer}/IRAC coverage and removes sources affected by bright stars or artefacts. We use this flag in the final stage of our analysis (Section \ref{sec:application}) when constructing our final prior list. For the prior list construction, we cross-match our master catalogue with the full \texttt{FARMER} dataset (964,506 sources) using a search radius of 0.3 arcsec. This results in a matched sample of 650,148 sources, which serves as the basis for our star-galaxy classification.

\begin{figure}
\centering
\includegraphics[width=0.98\columnwidth]{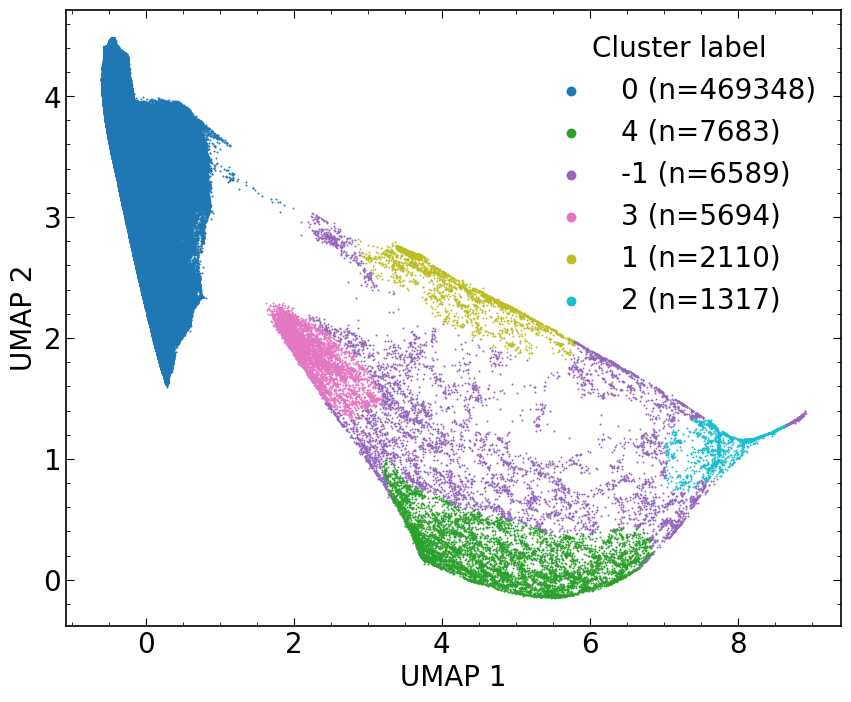}
\caption{A projection of sources onto the first two dimensions from the Uniform Manifold Approximation and Projection (UMAP) algorithm. These UMAP axes are abstract dimensions derived from the input photometry. Colours represent classifications by HDBSCAN. The main population of galaxies is shown in blue (Cluster 0). The other colours represent distinct populations of stellar sources identified by the clustering algorithm.}
\label{fig:umap_hsc}
\end{figure}

\subsection{Star-Galaxy Classification}
A significant challenge in constructing a prior list from deep optical/IR catalogues is the presence of stellar sources. Since most stars are not detectable radio sources (e.g \citealt{Driessen2024,Okwudili2025}), their inclusion in the prior list would introduce a population of non-emitting sources. This could cause the deblending algorithm to assign radio flux incorrectly from nearby galaxies to these stellar positions, thereby contaminating the measurements of genuine radio galaxies. Therefore, a reliable method for star-galaxy separation is an important preliminary step. 

\subsubsection{Methodology}
Our star-galaxy classification process begins with the          650,148 sources matched between our master catalogue and COSMOS2020. We use the HSC PDR3 photometry for classification because its improved depth enables more reliable morphological separation compared to earlier releases. We apply two selection criteria to ensure that we use robust photometry. First, we select sources with a signal-to-noise ratio in the K$_s$-band of SNR > 5, reducing the sample to 506,136 sources.

To classify these sources, we use a two-stage unsupervised machine learning process, following \cite{Cook2024}.  Unsupervised learning has the advantage of not being biased by a given training set, which is critical for faint source populations where complete and representative training data are unavailable \cite[see][for a review]{Fotopoulou2024}. First, we construct a high-dimensional feature space from the photometric data. We use magnitudes from the optical HSC survey and the near-infrared from VISTA to generate 28 colour combinations. These colours, combined with the 8 original magnitude measurements ($g, r, i, z, Y, J, H, K_s$), serve as 36 features for each source. While some colour features are mathematically redundant, we include all of them, as dimensionality reduction techniques like  Uniform Manifold Approximation and Projection (UMAP; \citealp{McInnes2020}) effectively extracts the most important underlying information from complete feature sets \citep{Cook2024}. Furthermore, we verified that a reduced set of independent features (e.g., a single magnitude band plus n-1 independent colours) produced a consistent classification, confirming the robustness of the approach. This high-dimensional feature space provides a more robust basis for distinguishing between stellar and galactic populations than is possible in any single two-dimensional parameter space (e.g \citealt{Strateva2001, Hewett2006}).

Before applying the machine learning algorithms, the feature set is first cleaned by removing sources with incomplete photometry, which yields a working sample of 492,741 sources. This feature set is then scaled using the \texttt{StandardScaler} package from \texttt{scikit-learn} \citep{Pedregosa2011}. This process applies a standard Z-score transformation, rescaling each feature to have a mean of zero and a standard deviation of one, which ensures that all features contribute equally to the analysis.

The first stage of the process applies the UMAP algorithm to this feature space \citep{McInnes2020}. UMAP is a non-linear dimensionality reduction technique that reduces the 36-dimensional feature space to a 10-dimensional embedding\footnote{These 10 output dimensions are abstract representations; each is a complex, non-linear combination of the original 36 photometric features and does not correspond directly to any single physical property like a specific colour or magnitude.} while preserving the essential topological structure of the data \citep{Becht2019}. Its ability to capture the complex, non-linear relationships inherent in multi-band photometric data makes it a superior choice to simpler linear methods like Principal Component Analysis (PCA; \citealt{Jolliffe2016}). Furthermore, UMAP has been shown to be more scalable and better at preserving the global structure of the data, crucial for separating the main star and galaxy clusters, than comparable techniques \citep{McInnes2020}.

The second stage applies the Hierarchical Density-Based Spatial Clustering of Applications with Noise (HDBSCAN) algorithm \citep{McInnes2017} to identify distinct clusters in the UMAP-reduced space. HDBSCAN is a density-based algorithm that can find clusters of varying shapes and densities without requiring the number of clusters to be specified beforehand, making it effective at isolating the primary stellar and galactic populations (e.g.,\citealt{Logan2020, Fotopoulou2024}).

\begin{figure}
\centering
\includegraphics[width=\columnwidth]{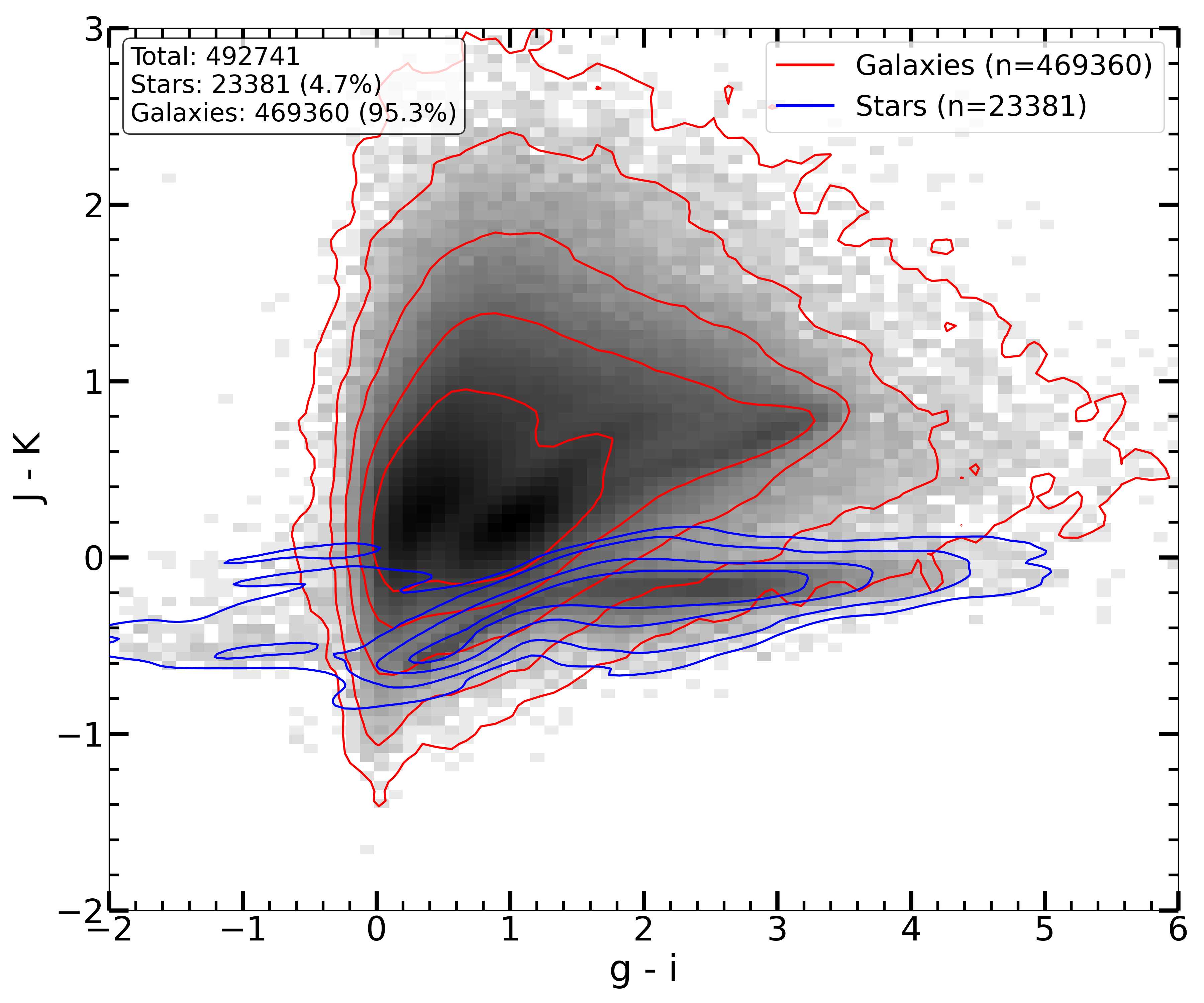}
\caption{$J-K$ vs. $g-i$ colour-colour diagram for the 492,741 sources. The underlying greyscale density map shows the distribution of the full sample. The overlaid contours trace the density of sources classified as galaxies (red) and stars (blue), illustrating the separation achieved by the high-dimensional analysis.}
\label{fig:cc_consolidated}
\end{figure}

\subsubsection{Application}
We applied this methodology to the filtered sample of $492,741$ sources derived from the cross-matched dataset. The process successfully separates the sources, yielding a catalogue of 
$469,360$ high-confidence galaxies ($95.3$ per~cent) and $23,381$ stars (4.7 per~cent), as detailed in Table \ref{tab:sg_breakdown}. The resulting UMAP projection, which clearly shows the distinct clusters, is presented in Figure~\ref{fig:umap_hsc}. The classification is visualized on the traditional $J-K$ vs. $g-i$ colour-colour diagram \citep[e.g.,][]{Jarvis2013} in Figure~\ref{fig:cc_consolidated}. In this two-dimensional projection, while the bulk of the populations lie on distinct loci, the contours reveal a region of degeneracy where stars and galaxies overlap. Traditional methods, which rely on defining a fixed boundary or simple curves relative to the stellar locus \citep[e.g.,][]{Jarvis2013}, inevitably suffer from contamination or incompleteness in this overlapping regime.  However, our machine learning approach uses the full high-dimensional feature space to resolve these degeneracies and successfully disentangles stars and galaxies that are indistinguishable in the $J-K$ vs. $g-i$ plane alone.

\begin{table}
\centering
\caption{Star-galaxy classification breakdown for the VISTA + HSC PDR3 dataset.}
\label{tab:sg_breakdown}
\begin{tabular}{lrr}
\toprule
\textbf{Classification} & \textbf{Number of Sources} & \textbf{Percentage} \\
\midrule
Galaxies & 469,360 & 95.3 per~cent \\
Stars & 23,381 & 4.7 per~cent \\
\midrule
\textbf{Final Total} & \textbf{492,741} & \textbf{100 per~cent} \\
\bottomrule
\end{tabular}
\end{table}

\begin{figure}
    \centering
    \includegraphics[width=\columnwidth]{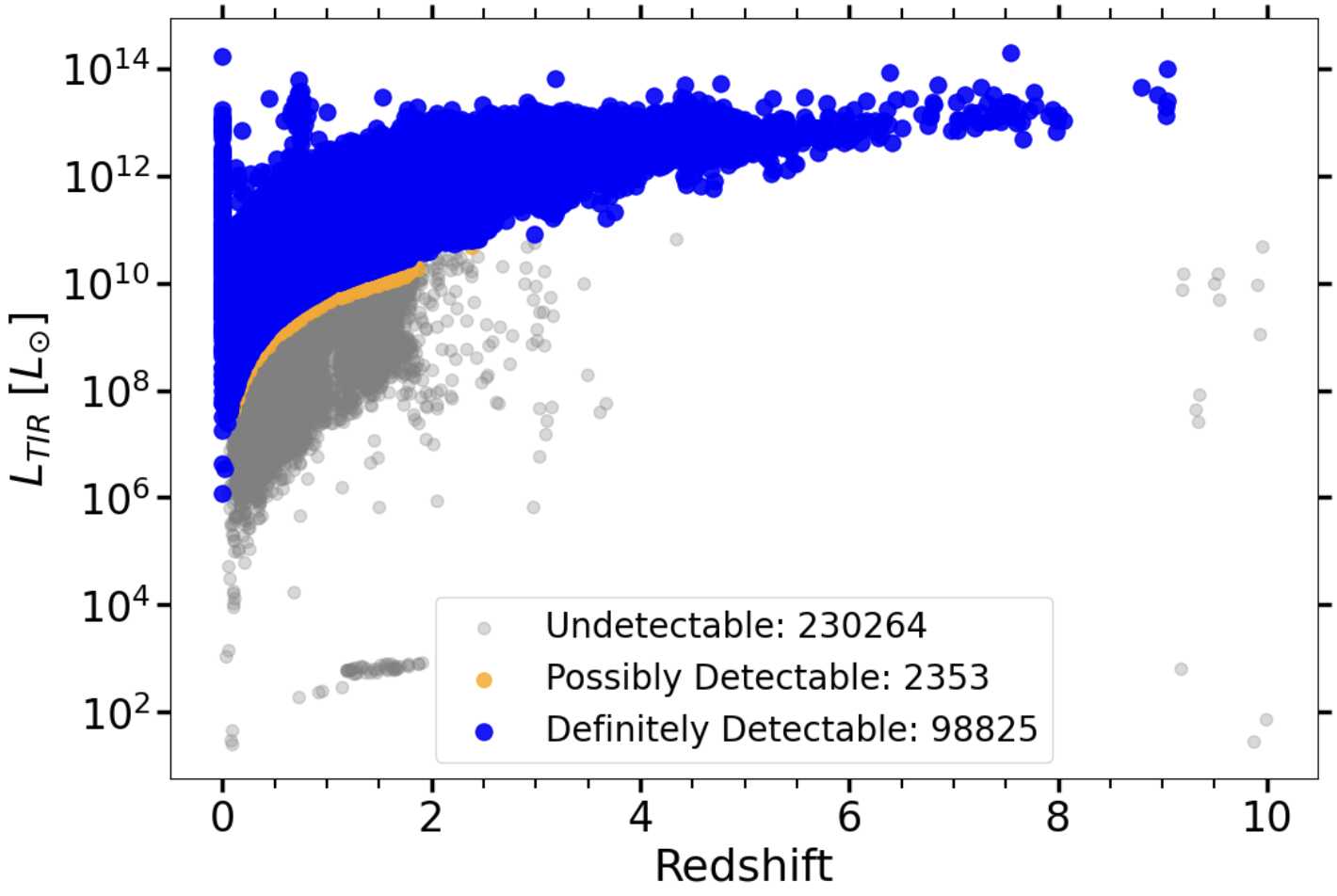}
    \caption{The distribution of the galaxy sample in the total infrared luminosity ($L_{\mathrm{IR}}$) versus redshift plane. Sources are classified based on their predicted 1.4\,GHz flux density relative to the MIGHTEE sensitivity limit. Blue dots represent sources classified as `Definitely Detectable' ($S_{\rm pred} > 1\,\mu$Jy), orange dots indicate `Possibly Detectable' sources (where the upper $L_{\mathrm{IR}}$ uncertainty bound implies $S_{\rm pred} > 1\,\mu$Jy), and grey dots show `Undetectable' sources.}
    \label{fig:radio_prediction} \end{figure} 

\subsection{The Galaxy Sample and Prior List Construction}
\label{sec:prior_construction}
Following star-galaxy classification, we define a galaxy sample from which deblending priors are constructed. We select all galaxies from our filtered sample that have a total infrared luminosity ($L_{\mathrm{IR}}$) estimate from the COSMOS2020 catalogue. This results in a galaxy list of 331,442 galaxies within the region of overlapping deep multi-wavelength coverage. This sample forms the basis for all subsequent prior list construction.

Constructing a prior list presents a trade-off between completeness and purity. An inclusive prior ensures that all potential radio-emitting galaxies are considered, but at the risk of including a large number of undetectable radio sources. As demonstrated by \citet{Hurley2017}, an excess of such priors can degrade the flux measurements of genuine detections by incorrectly partitioning the map's flux density. To investigate this trade-off and determine the optimal strategy for MIGHTEE, we construct two distinct prior lists from our galaxy sample, each designed to test a different deblending philosophy.

\subsubsection{\rev{\texttt{Comprehensive Prior}}}

The first prior list is designed to be as inclusive as possible. It incorporates all 331,442 sources from our galaxy sample that have an $L_{\mathrm{IR}}$ measurement. We supplement this list with 4,754  sources detected above 3$\sigma$ in the VLA-COSMOS 3\,GHz survey \citep{Smolcic2017}, as extracted by \citet{Jin2018}, specifically adding those that do not have counterparts in our optical/NIR sample. This results in a candidate list of 336,196 unique sources.
This \texttt{Comprehensive} L\textsubscript{IR} Prior represents a compromise, aiming for high completeness by including any source with evidence of significant star formation or AGN activity (via $L_{\mathrm{IR}}$) without being as exhaustive as including all $\sim$470,000 galaxies in the master sample.

\subsubsection{\rev{\texttt{The Radio-Likely Prior}}}

The second prior list is designed to be an optimized list containing only sources with a high probability of being detectable in the radio. In this context, we define `detectable' not as a strict $>5\sigma$, but as a source having a flux density sufficient to be statistically measurable. This \texttt{Radio-Likely Prior} is designed as a high-purity sample, containing sources with a high probability of detection while excluding those expected to be undetectable. This provides a cleaner and more effective input for XID+. The construction is a two-step process.

First, we predict the 1.4\,GHz radio flux density for each of the 331,442 galaxies in our galaxy sample based on their estimated infrared luminosity, using the well-established IRRC \citep{Helou1985,Condon1992,Yun2001}. The methodology is as follows:
\begin{enumerate}
    \item The rest-frame 1.4\,GHz radio luminosity ($L_{1.4}$) is calculated from the total infrared luminosity ($L_{\mathrm{IR}}$) using the linear correlation from \cite{CookRHW2024}, given by $\log_{10}(L_{\mathrm{IR}}/L_{\odot}) = 11.170 + 0.921 \log_{10}(L_{1.4} / 5\times10^{22}\,\mathrm{W\,Hz^{-1}})$.
    \item This radio luminosity is converted to an expected observed flux density ($S_{1.4}$) using the source's redshift ($z$) and corresponding luminosity distance ($D_L$), including a k-correction of the form $(1+z)^{1-\alpha}$. We assume a spectral index $\alpha=0.7$, a typical value for the synchrotron-dominated emission of star-forming galaxies at these frequencies \citep{Condon1992}.
    \item  Sources are then classified based on a comparison to the MIGHTEE map sensitivity. Those with a predicted flux density above our threshold (of $1\,\mu$Jy) are flagged as `Definitely Detectable'. To account for uncertainties, we also identify sources that fall below this threshold but whose predicted flux density based on the upper error bound of their $L_{\mathrm{IR}}$ exceeds it; these are flagged as `Possibly Detectable'.
\end{enumerate}

The results of this prediction step are shown in Figure \ref{fig:radio_prediction}. As expected, the 98,825 sources predicted to be `Definitely Detectable' occupy the high-$L_{\mathrm{IR}}$ regime across all redshifts, with the 2,353 `Possibly Detectable' sources forming a boundary region where detection is plausible given $L_{\mathrm{IR}}$ uncertainties. \rev{We note that the tail of `Definitely Detectable' sources extending to very high redshift in Figure~\ref{fig:radio_prediction} should be interpreted with caution due to the lack of FIR constraints in the SED fitting. Furthermore, physical effects, such as increased inverse-Compton cooling off the Cosmic Microwave Background (CMB) photons, are expected to suppress the observed synchrotron emission at these redshifts \citep[e.g.,][]{Murphy2009, Lacki_Thompson2010, Wu2017, Whittam2025}. However, these potentially spurious high-redshift sources constitute a negligible fraction ($<0.1\%$) of the full prior list and therefore do not impact the derived source counts or deblending results.}

Second, this core list of likely radio-bright sources is supplemented with the same VLA radio sources used for the comprehensive prior to ensure completeness for known radio-emitting sources. This results in a candidate list of 106,164 unique galaxies for the \texttt{Radio-Likely Prior}.

\subsubsection{\rev{Validation of Prior Selection}}
\label{sec:prior_validation}

\rev{To validate the reliability of our SED-derived $L_{\rm IR}$ selection, particularly given the potential uncertainties in the energy-balance estimates which lack direct FIR constraints, we compare the IRRC-selected subset of our prior list to the updated Super-deblended catalogue \citep{Jin2018, Sillassen2024}. This catalogue serves as a valuable independent reference because its prior list was constructed using a different approach. While our method selects sources based on their predicted radio flux derived from $L_{\rm IR}$, the Super-deblended prior list (containing $\sim$200,000 sources) is a stellar mass-limited sample. It was constructed by selecting sources from the COSMOS2020 catalogue \citep{Weaver2022}, the VLA 3 GHz survey \citep{Smolcic2017}, and the A3COSMOS catalogue \citep{Liu2019}. Since stellar mass correlates strongly with star formation rate \citep[e.g.,][]{Noeske2007, Whitaker2012}, and radio luminosity is a direct tracer of star formation, our SED-derived $L_{\rm IR}$ prior and the \citet{Jin2018} stellar-mass limited approach should target essentially the same underlying galaxy population. Any significant differences would therefore indicate potential issues with our prior list fidelity.}

\rev{We cross-match our prior list with the Super-deblended catalogue using a search radius of 0.3 arcsec. We restrict this analysis to the clean area defined by the \texttt{FLAG\_COMBINED} flag and the stellar mask (detailed in Section \ref{sec:application}). This filtering reduces our galaxy sample to 298,894 sources. Within this area, our IRRC-based method identifies a \texttt{Radio-Likely Prior} subset of 87,259 sources. When matched against the super-deblended catalogue, we find that $>98$ per cent of our sources have counterparts in the independent catalogue. This near-perfect agreement with a high-fidelity benchmark confirms that our $L_{\rm IR}$-based selection effectively isolates the detectable radio population that follows the IRRC. Furthermore, it demonstrates that the uncertainties in $L_{\rm IR}$ arising from the lack of FIR data in the SED fitting do not significantly compromise the selection of the radio-emitting sample.}

\rev{Having validated the high purity of the \texttt{Radio-Likely Prior} against independent data, we now turn to simulations to rigorously test the deblending accuracy. The relative performance of these two prior lists, one designed for inclusivity and the other for purity, will be rigorously tested in Section \ref{sec:sim_validation} to determine the optimal strategy for our final scientific analysis.}

\section{XID+ probabilistic deblending technique}
\label{sec:xidplus}
Having established a prior list of source positions from ancillary data, we now address the challenge of measuring their flux densities in the confused MIGHTEE radio maps. We use the probabilistic deblending framework XID+ \citep{Hurley2017} to perform this task. Even among prior-driven deblending approaches, many rely on maximum-likelihood estimation to assign flux (e.g., \texttt{FASTPHOT}, \citealt{Bethermin2010}). While computationally efficient, these approaches can struggle in highly confused regimes, often incorrectly assigning the total flux density of a blend to a single source and failing to provide robust uncertainty estimates (e.g.,\citealt{Roseboom2010,Wang2014}).

XID+ overcomes these limitations by adopting a fully probabilistic Bayesian framework (\citealt{Hurley2017}). Instead of seeking a single best-fit solution, XID+ explores the entire parameter space to map out the full posterior probability distribution for the flux of every source \citep{Hurley2017}. This approach has several advantages. First, it provides uncertainty quantification by naturally capturing the credible intervals and degeneracies that arise when multiple sources are close together. Second, the joint posterior reveals covariance between sources, quantifying how the flux density estimate of one source is correlated with its neighbours. Third, the Bayesian framework correctly models the behavior of sources at or below the noise level.

The core assumption of XID+ is that the observed map data, $d$, can be modelled as a linear combination of the flux densities, $S$, of a known set of $N_s$ prior sources, each treated as a point source convolved with the instrument's Point Spread Function (PSF), plus a background ($B$) and noise component ($n$). For this work, we use the Gaussian restoring beam from the low-resolution MIGHTEE data. This noise is modelled as a Gaussian that accounts for both thermal noise and confusion noise arising from the blending of faint unresolved sources (e.g.,\citealt{Condon1974}). For a map flattened into a vector of $M$ pixels, this can be expressed as:
\begin{equation}
    d = PS + n + B,
    \label{eq:xid_model}
\end{equation}
where: $P$ is the $M \times N_s$ pointing matrix, which represents the PSF response of each prior source. It is constructed by placing a model of the instrument PSF at the sky position of each source in the prior list.

The Bayesian paradigm seeks the posterior probability of the fluxes given the data, $p(S|d)$, which is given by Bayes' theorem:
\begin{equation}
    p(S|d) \propto p(d|S) \times p(S).
    \label{eq:bayes}
\end{equation}
Here, $p(d|S)$ is the likelihood of observing the data given a set of fluxes, and $p(S)$ is the prior probability distribution of the fluxes. The likelihood is modelled as a multivariate Gaussian distribution based on the noise properties of the map. The standard XID+ framework calculates the total variance by adding the thermal variance (from a user-provided map) and a fitted confusion variance in quadrature ($\sigma_{\mathrm Tot} ^2 = \sigma^2 + \sigma_\mathrm{conf}^2$). However, a pure thermal noise map was not available for the MIGHTEE DR1 data products; instead, a total noise RMS map, which already accounts for both effects, was provided \citep{Hale2025}. This means the standard likelihood calculation within XID+ could not be used without modification.

For the deblending performed in this work, we use a simple, uninformative (flat) prior for the flux of each source, typically bounded between 0 and a sufficiently high value, to avoid biasing the resulting measurements towards any pre-conceived model \citep{Hurley2017, Wang2024}. The role of the ancillary data is therefore to define an optimal and pure list of source positions, rather than to inform the flux priors directly.

Solving for the full posterior distribution in Equation \ref{eq:bayes} is a high-dimensional problem that is computationally intractable to solve analytically. XID+ addresses this by using the probabilistic programming language \texttt{Stan} \citep{Carpenter2017}. Specifically, it uses the No-U-Turn Sampler (NUTS, \citealt{Hoffman2014}), to efficiently draw samples from the high-dimensional posterior distribution. To accommodate the MIGHTEE DR1 noise map, we modified the XID+ Stan code to use the pre-computed total variance from the DR1 RMS map at each pixel ($\sigma_{\rm Tot}^2 = \sigma^2_{\mathrm{DR1\_RMS}}$) directly in the likelihood calculation. While this means we do not derive a separate confusion noise component, it ensures the deblending is performed using the most accurate, position-dependent noise estimate available. An important parameter governing the sampler's exploration is \texttt{max\_treedepth}, which controls the complexity of the integration paths, with a default value of 10. Testing on real MIGHTEE data in Section \ref{sec:real_flux_recovery} shows that increasing this to 15 was required for convergence. Following the recommendation of \cite{Hurley2017}, we run four independent MCMC chains to ensure convergence.

Fitting the whole map at once is computationally unfeasible, so the analysis is partitioned into small tiles using the HEALPix framework \citep{Gorski2005}, each $\sim$ 3.4\,arcmin across (HEALPix order 10). However, to ensure an accurate flux measurement for sources within one of these small tiles, the deblending fit must account for flux spilling in from bright sources in the surrounding area. Therefore, the fit is performed on a much larger `fitting' tile, $\sim$ 27.5\,arcmin across (HEALPix order 7), which fully encompasses the smaller results tile. Once the fit on the {larger tile} is complete, only the results for sources within the central small tile are retained. The primary output for each source within a successfully fitted tile is a chain of samples from its posterior distribution. The point estimates for the deblended flux densities for each source are taken as the median (50th percentile) of its posterior distribution for each source. The final catalogue is produced by repeating this process for all tiles across the map and combining the results.

\section{Deblending Validation with Simulations}
\label{sec:sim_validation}
While XID+ has been extensively and successfully applied to confused FIR and sub-millimetre data from instruments like \textit{Herschel} and SCUBA-2 (e.g., \citealt{Hurley2017, Pearson2017, Shirley2021, Wang2024}), its application to deep radio continuum data is less established. Radio surveys present a unique set of challenges, including more complex source morphologies (e.g., core-jet-lobe structures in AGNs; \citealt{ Mingo2019,Hardcastle2020}), which violates the point-source assumption of XID+ (see Section \ref{sec:xidplus}). 

Therefore, before applying XID+ to the MIGHTEE data, it is essential to conduct a thorough validation using realistic simulations. These simulations provide a known ground truth, allowing us to test the performance of XID+ under radio-specific conditions and to identify the optimal configuration for our deblending strategy.

\begin{figure*}
    \centering
    \begin{subfigure}[b]{0.24\textwidth}
        \includegraphics[width=\linewidth]{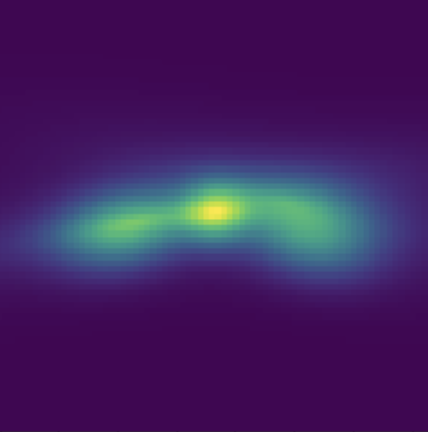} 
    \end{subfigure}
    \hfill
    \begin{subfigure}[b]{0.24\textwidth}
        \includegraphics[width=\linewidth]{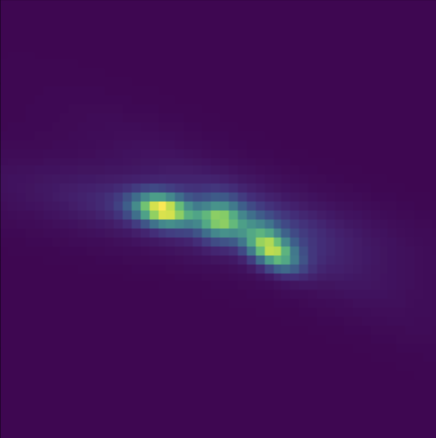} 
    \end{subfigure}
    \hfill
    \begin{subfigure}[b]{0.24\textwidth}
        \includegraphics[width=\linewidth]{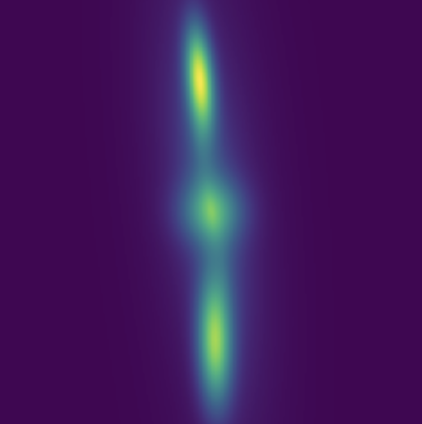} 
    \end{subfigure}
    \hfill
    \begin{subfigure}[b]{0.24\textwidth}
        \includegraphics[width=\linewidth]{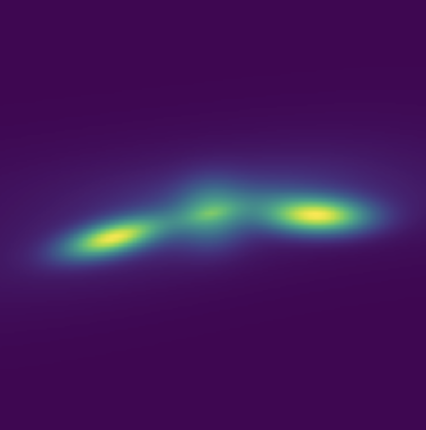} 
    \end{subfigure}

    \vspace{0.2cm} 

    \begin{subfigure}[b]{0.24\textwidth}
        \includegraphics[width=\linewidth]{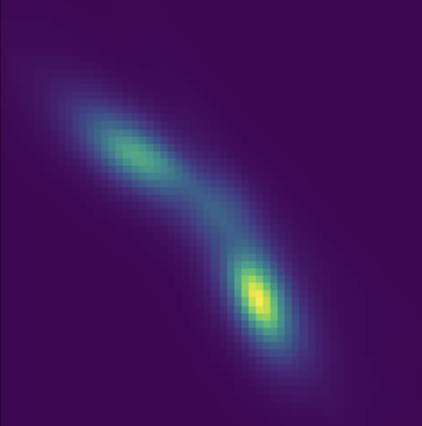} 
    \end{subfigure}
    \hfill
    \begin{subfigure}[b]{0.24\textwidth}
        \includegraphics[width=\linewidth]{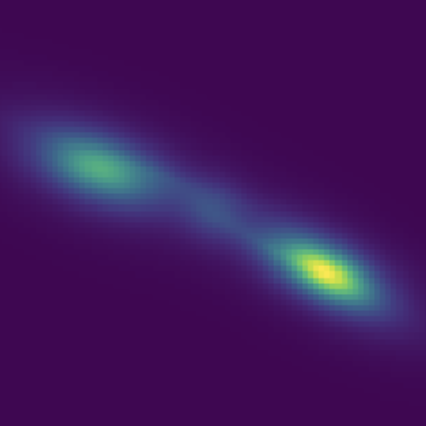} 
    \end{subfigure}
    \hfill
    \begin{subfigure}[b]{0.24\textwidth}
        \includegraphics[width=\linewidth]{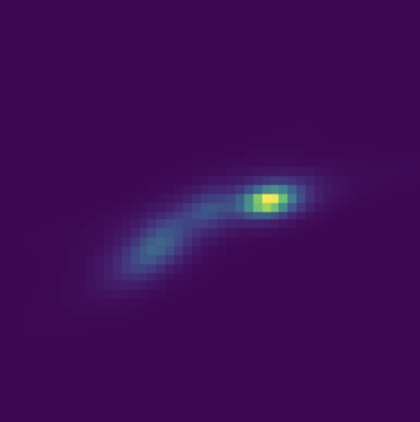} 
    \end{subfigure}
    \hfill
    \begin{subfigure}[b]{0.24\textwidth}
        \includegraphics[width=\linewidth]{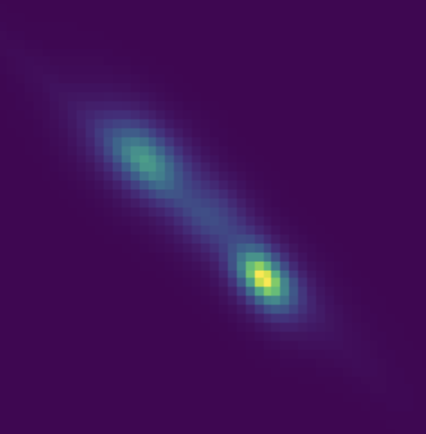} 
    \end{subfigure}

    \caption{Examples of the complex, multi-component AGN morphologies generated by our simulation pipeline. The top row displays four Fanaroff-Riley Type I (FRI) sources, modeled with bent jets. The bottom row displays four Fanaroff-Riley Type II (FRII) sources, featuring distinct lobes and hotspots. All images are $50 \times 50$ pixels in size, with a pixel scale of 1.1 arcsec.}
    
    \label{fig:sim_morphologies}
\end{figure*}

\subsection{The Simulation Framework}
\label{sec:sim_framework}
Our simulations generate realistic radio sky models that capture the morphological diversity of galaxies and the basic properties of MIGHTEE. The framework, adapted from \citet{Harrison2020} and implemented in a custom Python script, uses the \texttt{galsim} \citep{Rowe2015} software package to draw galaxy models into a simulated map. The properties of the source population are drawn from the Tiered Radio Extragalactic Continuum Simulation (T-RECS) catalogue \citep{Bonaldi2019,Bonaldi2023}.

The modelling of galaxy morphology is tailored to the source type to achieve a high degree of realism:
\begin{itemize}
    \item \textbf{SFGs:} These are modeled using Sersic profiles \citep{Sersic1963}. The Sersic index ($n$) is set based on the galaxy's optical classification (e.g., $n=1$ for disk-like, $n=4$ for bulge-like), while the scale radius, flux, and orientation are taken from the T-RECS catalogue values.
     \item \textbf{AGNs:} AGN modelling depends on their radio classification. Compact AGNs (FSRQs, BL Lacs) are modeled as single, compact Gaussian profiles. 
     
     The more complex Steep-Spectrum AGNs (SS-AGNs), corresponding to Fanaroff-Riley Class I (FRI) and II (FRII) sources  \citep{Fanaroff&Riley1974}, are modeled with a multi-component structure. The total flux is distributed among a core, jets, and lobes, with the fractional contribution of each component depending on the \texttt{Rs} parameter, defined as the ratio between the total projected source size and the projected distance between the two bright hot spots, from the T-RECS catalogue, which distinguishes between core- and lobe-dominated structures. The overall source size is similarly apportioned between a central region and outer structures, with the relative extent of the jets and lobes also governed by the \texttt{Rs} parameter to model the brighter, more extended jets of FRIs and the dominant lobes of FRIIs. 
     
     To mimic the asymmetric structures observed in real radio surveys (e.g., \citealt{Laing2002}), we implement a bending model where each side of the AGN is assigned randomized parameters for an angular offset, a shear, and a positional shift. The distributions from which these random values are drawn are different for FRI and FRII sources, allowing for the larger, more distorted morphologies characteristic of FRI jets. The final model components are then: a compact Gaussian core; composite jets made of a central, straight exponential profile and two outer bent exponential profiles; and lobes modeled as either diffuse Gaussians (for FRIs) or sharper Moffat profiles to represent hotspots (for FRIIs), reflecting the different termination shocks in the two classes (e.g., \citealt{Hardcastle2020}). Each component's final position and orientation are determined by the bending model, ensuring a morphologically diverse and realistic AGN population. Figure \ref{fig:sim_morphologies} demonstrates the morphological complexity incorporated into our simulations. The top row displays FRI sources, characterized by their core-dominated, bent-jet structures, while the bottom row shows FRII sources with distinct hotspots and lobes. While these parametric models do not capture the full physical evolution inherent in hydrodynamic simulations \citep[e.g.,][]{Giri2024} \rev{or the full morphological diversity revealed by the sensitivity of MeerKAT, this level of realism is sufficient for our purposes. The brightest, most complex sources are masked in our analysis, meaning XID+ is primarily tasked with deblending the fainter, likely to be more compact population, which is well-represented by our simulation.}
\end{itemize}

The primary validation tests were conducted on a simulated map configured to match the MIGHTEE low-resolution data, with a pixel scale of 1.1 arcsec and where the source models are convolved with a Gaussian PSF of 8.6\,arcsec FWHM. The simulation includes sources from the T-RECS catalogue down to a flux limit of 0.5 $\mu$Jy, sources fainter than this limit contribute negligibly to the final map statistics given the target noise level. For these tests, we added uniform Gaussian noise with a $\sigma$ of 1.3~$\mu$Jy/beam, a value representative of the thermal noise in the deepest parts of the MIGHTEE fields \citep{Hale2025}. 

\begin{figure*}
    \centering
    \includegraphics[width=1.95\columnwidth]{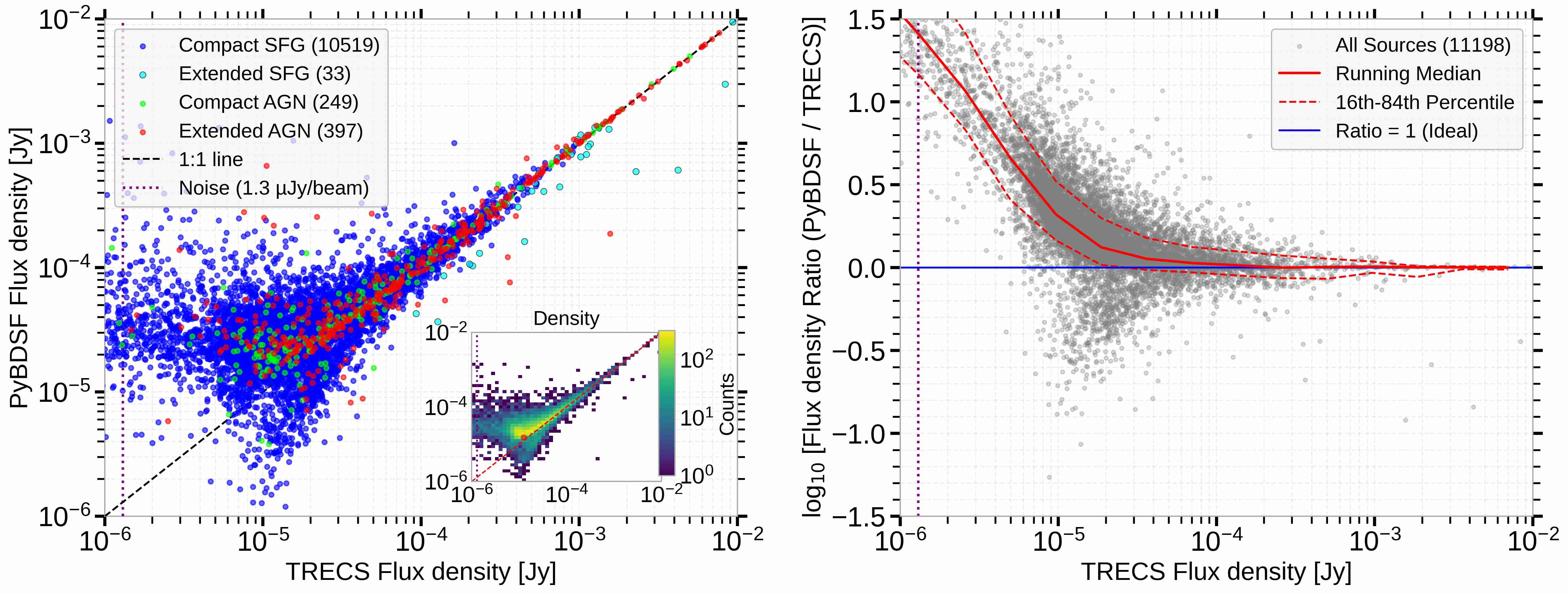}
    \caption{Left: Comparison of input T-RECS catalogue flux densities versus flux densities recovered by a multi-stage PyBDSF source-finding process on the simulated map. Sources are coloured by type: Compact SFGs (blue), Extended SFGs (cyan), Compact AGNs (green), and Extended AGNs (red). The dashed black line indicates a perfect 1:1 correlation. \rev{The inset shows a 2D density histogram of the points, with darker shades indicating higher source density.} Right: The log-ratio of recovered PyBDSF-to-true flux versus true flux for all sources (grey points). The solid red line is the running median of this ratio, with the dashed red lines showing the 16th-84th percentile range. The solid blue line indicates a perfect recovery ratio of 1.}
    \label{fig:pybdsf_bias}
\end{figure*}

\subsection{Validating the Simulation}
\label{sec:sim_validation_and_motivation}
Before using the simulation to test our deblending approach, we must first validate that the simulation itself is a reliable representation of the radio sky. 

For a quantitative test, we apply a blind source-finding algorithm Python Blob Detection and Source Finder (PyBDSF; \citealt{Mohan2015}) to the simulated map. We use a multi-stage process similar to that for the official MIGHTEE DR1 catalogues (\citealt{Hale2025}) as well as for the ELAIS-N1 field from the LOFAR Two-metre Sky Survey Deep Fields Data Release 2 \citep{Shimwell2025}. This approach is necessary for deeply confused fields like those studied here as a single-pass source-finding run is insufficient because the presence of numerous faint, undetected sources elevates the measured total noise ($\sigma_{\rm Tot}$). This artificially high noise level can prevent the detection of other faint sources that would otherwise meet the detection threshold.

The multi-stage approach consists of three main steps. For all three steps, we run PyBDSF using its default parameters, which include a 3$\sigma_{\rm Tot}$ island threshold and a 5$\sigma_{\rm Tot}$ detection threshold. First, an initial pass of PyBDSF is run on the original map. This initial run generates both a catalogue of the sources and a corresponding residual map. Secondly, we run PyBDSF on the residual map and we obtain a more accurate map of the background total noise. In the final step, this improved noise map is used as input for a second PyBDSF run on the original. This procedure allows the source-finder to identify faint sources, leading to a more complete and accurate source catalogue.
 
The results from this process are shown in Figure \ref{fig:pybdsf_bias}, which we use to validate our simulation. At bright flux densities ($S_{1.4} > 100\mu$Jy), the recovered flux densities are in good agreement with the input values, lying tightly along the one-to-one correlation line. This holds true for all source types, including the morphologically complex Extended AGNs and Extended SFGs.

However, at flux densities below $S_{1.4} \sim 100 \mu$Jy, and particularly for the dominant population of Compact SFGs, a systematic overestimation of flux or "flux boosting" begins to emerge and becomes increasingly severe towards lower flux densities. This is a well-documented effect in deep radio surveys (e.g., \citealt{Jauncey1968}; \citealt{Zwart2014} and references within), and its impact on the MIGHTEE DR1 source counts was characterized and corrected for using simulations by \citet{Hale2025}. The right-hand panel quantifies this trend: the running median of the flux ratio, which is close to unity for bright sources, gradually climbs, reaching a factor of three (0.5 dex) at the faintest flux levels ($S_{1.4}  \sim 10 \mu$Jy). 

 The analysis shows that even after the multi-stage process mitigates noise-related biases, we still have a systematic overestimation of flux at faint levels. This is due to source confusion. When multiple physically distinct sources lie closer than the telescope's resolution, any source-finder will tend to group their emission into a single detection, incorrectly assigning their combined flux to one position. This demonstrates that while our multi-stage PyBDSF approach is valid for finding sources, it is not well-suited to providing accurate faint-source flux measurements in a confused map. For this task, a deblending algorithm like XID+ is required for probing the faint radio sky.

\begin{figure*}
    \centering
    \includegraphics[width=0.8\textwidth]{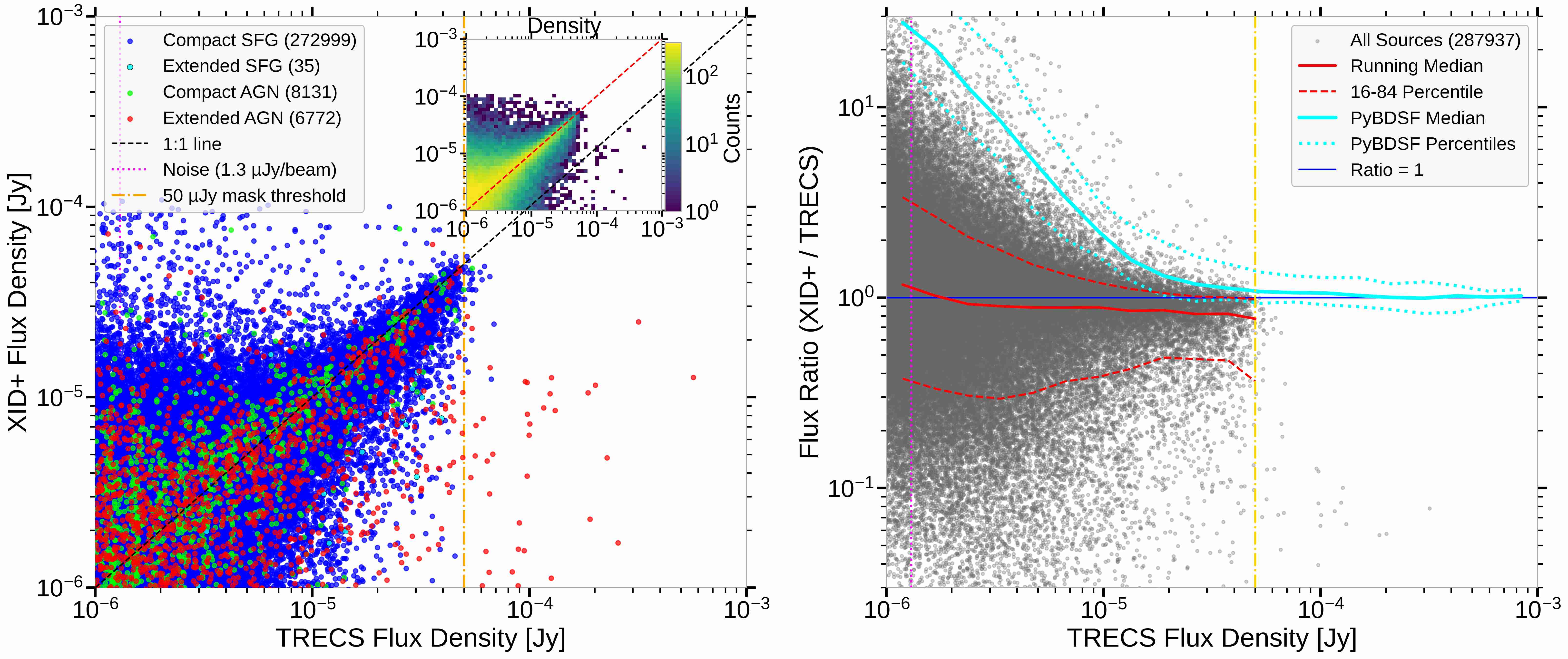}\\
    \includegraphics[width=0.8\textwidth]{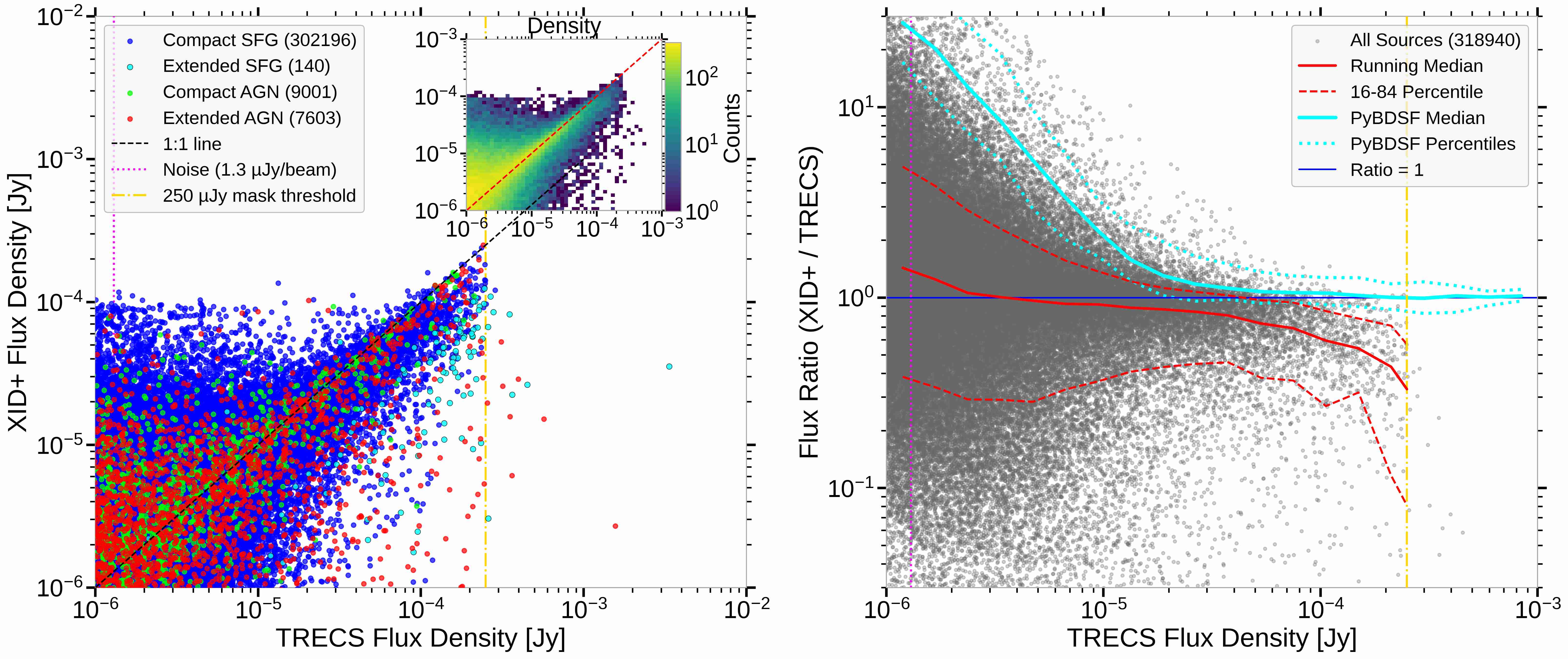}
    \caption{XID+ deblending performance using the \textbf{full T-RECS prior} with masks derived from the PyBDSF \textbf{SRL} catalogue. Each plot consists of two panels. The left panel shows the deblended flux from XID+ versus the true input flux from the T-RECS simulation, with sources coloured by type: Compact SFG (blue), Extended SFG (cyan), Compact AGN (green), and Extended AGN (red). The dashed black line indicates a perfect 1:1 recovery, and the inset shows a 2D density histogram of the points. The right panel shows the log-ratio of deblended-to-true flux versus true flux for all sources (grey points). The solid red line is the running median of this ratio, with the dashed red lines showing the 16th-84th percentile range. The solid cyan line shows the running median from PyBDSF, with dotted cyan lines showing its percentile range. The solid blue line indicates a perfect recovery ratio of 1. Top plot: masking all sources brighter than 50~$\mu$Jy using the SRL catalogue. Bottom Plot: masking all sources brighter than 250~$\mu$Jy using the SRL catalogue.}
    \label{fig:sim_flux_comp_full}
\end{figure*}

\subsection{Determining the Optimal Deblending Strategy}
\label{sec:xidplus_optimization}
We now use the simulation as a testbed to determine the most effective configuration for implementing XID+. We focus on three parameters: the mask type used for masking of bright, complex sources, the level threshold of the masking, and the composition of the prior input list.

\subsubsection{Masking Strategy: Threshold and Type}
\label{sec:masking_strategy}

The XID+ framework assumes point-like sources, an assumption that breaks down for the very bright, resolved, or morphologically complex objects that blind source-finders detect most easily. These sources must be masked to prevent them from corrupting the flux density estimates of neighbouring faint sources. 

Our masking is based on PyBDSF Gaussian models. We first select bright sources with a total or peak flux density above a certain threshold. We then construct a 2D model of the source using the fitted parameters (flux density, major axis, minor axis, position angle) from PyBDSF for each of these bright sources. The individual source mask is generated by identifying all pixels within this model that exceed a 1$\sigma_{\rm Tot}$ level relative to the total noise of the map. 

We determine the optimal flux density thresholds for masking by examining the performance of the blind source-finder (PyBDSF) in our simulations (Figure \ref{fig:pybdsf_bias}). We define two thresholds based on the regimes where blind extraction becomes unreliable due to confusion:

\begin{itemize}
    \item The $S>50$ $\mu$Jy mask: This mask removes all sources with flux densities $S_{1.4} > 50\,\mu$Jy ($\log_{10}(S) > -4.3$). As shown in the right panel of Figure \ref{fig:pybdsf_bias} (cyan line), this nominal threshold corresponds approximately to the onset of systematic flux boosting, where the PyBDSF recovered flux begins to systematically deviate from the input values. By masking sources brighter than this limit, we remove the population that is reasonably well-recovered by blind source finding, creating a "cleaner" map for XID+ to deblend the remaining faint, confused population where blind extraction fails.

    \item The $S>$ 250 $\mu$Jy mask: This mask removes only sources with $S_{1.4} > 250\,\mu$Jy ($\log_{10}(S) > -3.6$). This threshold corresponds to the bright regime where the PyBDSF recovery is unbiased (ratio $\approx 1$) and the scatter is minimal. This strategy preserves a larger area and a greater number of intermediate-brightness sources for deblending, providing a test of XID+ in a more challenging environment with higher source density.

\end{itemize} 

In addition to the flux density threshold, we test the impact of the catalogue type used to generate the mask. PyBDSF's process involves first identifying `islands' of contiguous emission above a given threshold, and then fitting one or more Gaussian components to model the flux within each island. From this, it produces two primary catalogues: a Gaussian component list (GAUL), which contains the parameters of every individual Gaussian component fitted to the emission islands, and a source list (SRL), which groups related Gaussian components into what the algorithm interprets as single astrophysical objects. These catalogues are known to trace different aspects of the radio source population, particularly for bright and extended sources. By testing these different configurations, we can determine which masking approach yields the most reliable deblended flux densities for the remaining faint source population.

\subsubsection{Prior List Selection}
The performance of XID+ is dependent on the quality of the prior catalogue. We test two scenarios:
\begin{itemize}
    \item \textit{Full T-RECS Prior:} A complete prior that is analogous to using the unfiltered optical/NIR catalogue for the real data. It contains the true positions of all sources from the simulation that fall within the unmasked regions. 
    \item \texttt{Radio-Likely Prior:} An optimized prior, the simulation analog of the \texttt{Radio-Likely Prior} from Section \ref{sec:prior_construction}, constructed by selecting all sources from the full T-RECS catalogue with an intrinsic flux greater than 1 $\mu$Jy. This tests how XID+ performs with a high-purity but necessarily incomplete prior list.
\end{itemize}
The number of sources in each of these sub-samples for the different masking configurations is detailed in Table \ref{tab:sim_source_counts}.

\begin{table}
    \centering
    \caption{Number of T-RECS sources in the simulated field under different masking and prior selection criteria.}
    \label{tab:sim_source_counts}
    \begin{tabular}{lrr}
        \toprule
        \textbf{Simulation configuration} & \multicolumn{2}{c}{\textbf{Number of Prior Sources}} \\
        & \textbf{Full Prior} & \textbf{\texttt{Radio-Likely Prior}} \\
        \midrule
        Total Sources in Field (no masked) & 329,178 & 80,574 \\
        \midrule
        \multicolumn{3}{l}{\textbf{Using PyBDSF GAUL-based Masking}} \\
        \quad Remaining after 50~$\mu$Jy Masking & 279,598 & 65,738 \\
        \quad Remaining after 250~$\mu$Jy Masking & 317,338 & 77,562 \\
        \midrule
        \multicolumn{3}{l}{\textbf{Using PyBDSF SRL-based Masking}} \\
        \quad Remaining after 50~$\mu$Jy Masking & 288,014 & 67,307 \\
        \quad Remaining after 250~$\mu$Jy Masking & 319,017 & 77,673 \\
        \bottomrule
    \end{tabular}
\end{table}

\begin{figure*}
    \centering
    \includegraphics[width=0.8\textwidth]{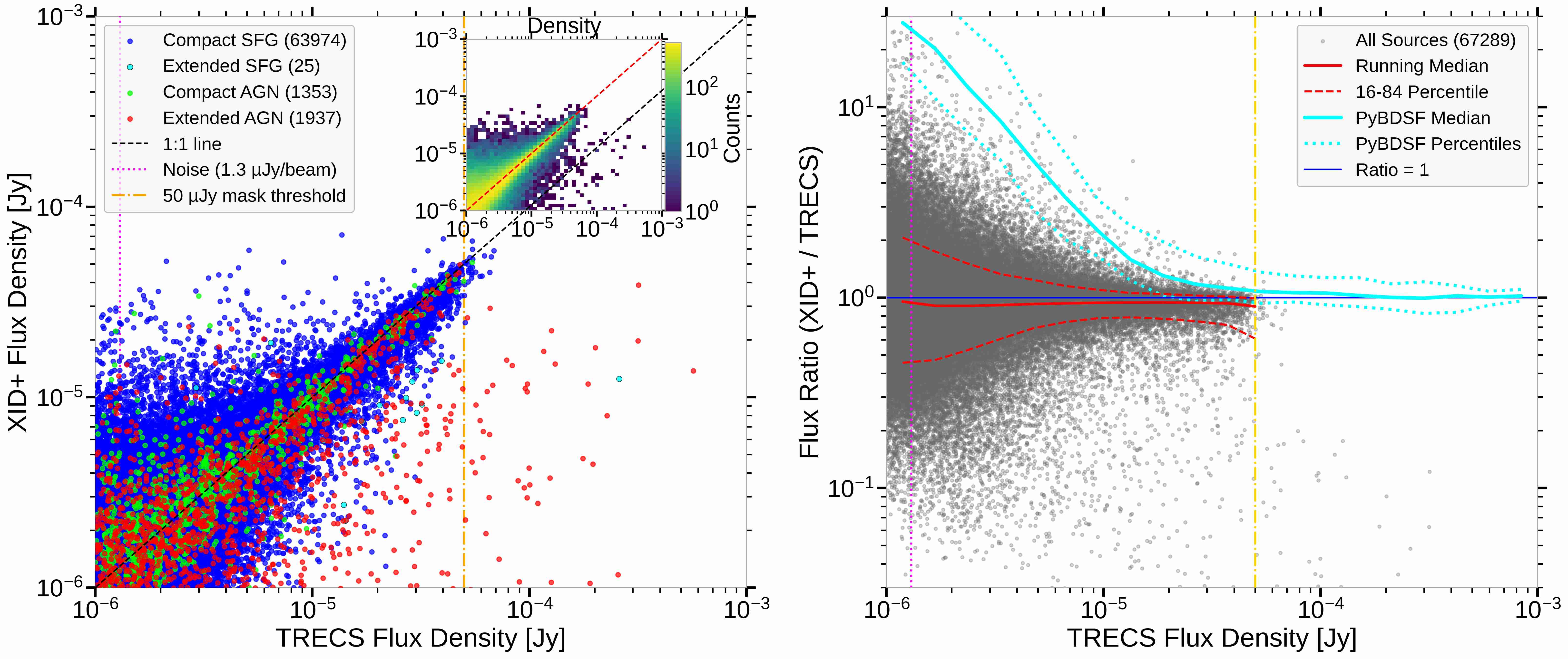}\\
    \includegraphics[width=0.8\textwidth]{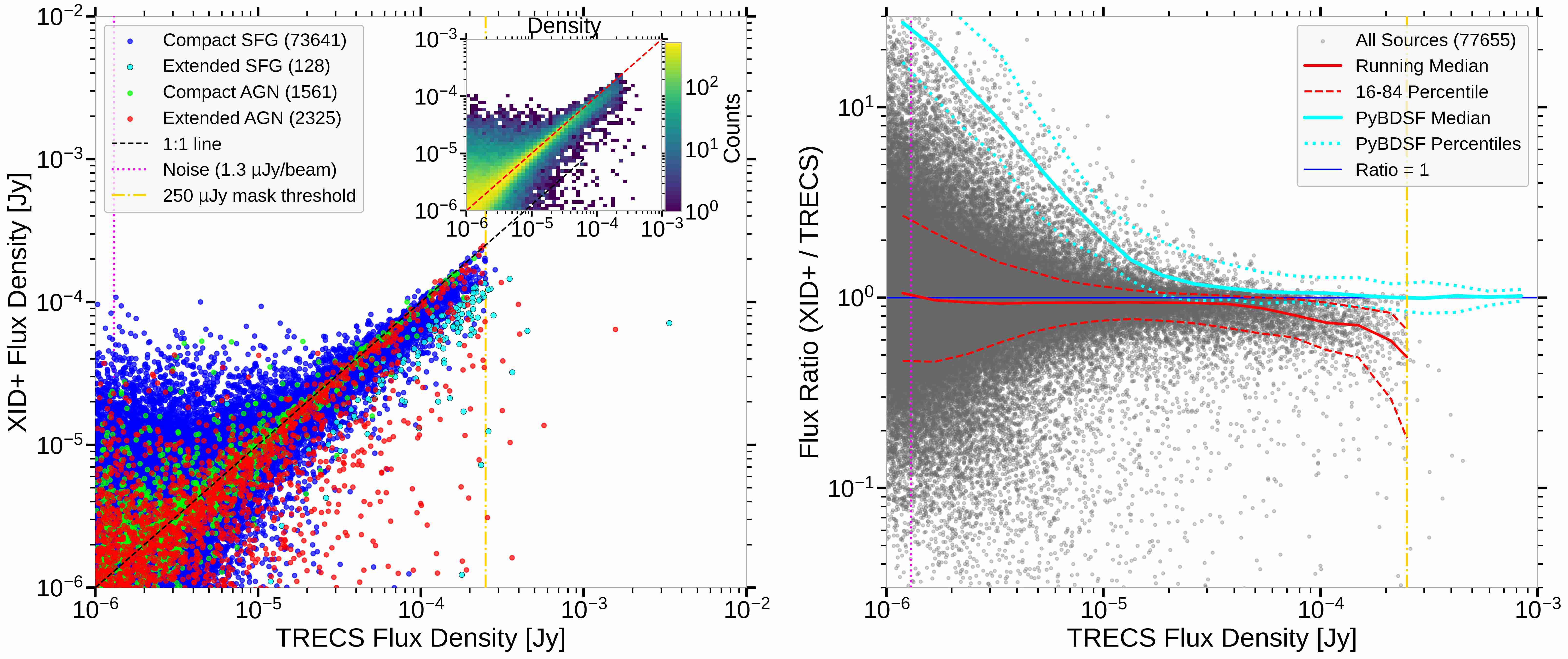}
    \caption{XID+ deblending performance using the optimal but incomplete \texttt{Radio-Likely Prior}  with masks derived from the PyBDSF \textbf{SRL} catalogue. The panels and symbols are the same as in Figure \ref{fig:sim_flux_comp_full}. Top: Performance when masking sources brighter than 50~$\mu$Jy using the GAUL catalogue. Bottom: Performance when masking sources brighter than 250~$\mu$Jy using the GAUL catalogue. For the sources included in the prior, the flux recovery is excellent in both cases, but the 50~$\mu$Jy mask yields a tighter correlation with lower scatter, producing the most reliable flux measurements.}
    \label{fig:sim_flux_comp_fir}
\end{figure*}

\subsection{Deblending Performance and Flux Recovery}
\label{sec:sim_flux_recovery}

We assess the performance of the deblending strategies by examining the accuracy (bias, via the running median of the flux ratio) and the precision (scatter, via the 16th-84th percentile range) of the deblended flux densities relative to the true T-RECS values. 

\subsubsection{Improvement Over Blind Source Finding}
PyBDSF suffers from strong flux boosting towards faint flux density (Figure \ref{fig:pybdsf_bias}). All XID+ deblending strategies show a marked improvement for the faint source population. They all maintain a running median flux ratio much closer to zero across the deblended flux range, with typical deviations of less than 0.05 dex for faint sources. Even the least optimal strategy, therefore, successfully addresses the primary bias inherent in blind source-finding methods. However, while the accuracy (bias) is consistently improved, the precision (scatter) of the deblended flux densities varies significantly among the different strategies.

\subsubsection{The Impact of Prior Catalogue Quality}
\label{sec:prior_impact}
The choice of prior list is the dominant factor affecting the precision of the deblended flux densities. When using the \texttt{Full Prior}, which contains all sources from the T-RECS simulation in the unmasked area, the deblending precision is relatively poor. As shown in Figure~\ref{fig:sim_flux_comp_full} (top; 50~$\mu$Jy mask), the 16th--84th percentile range of the flux ratio widens from approximately 0.4\,dex at 30\,$\mu$Jy to 0.8\,dex at 3\,$\mu$Jy. This indicates that a typical 3\,$\mu$Jy source has an uncertainty of almost an order of magnitude. 

This high scatter is a direct consequence of over-deblending, where the algorithm incorrectly partitions flux among the dense list of faint priors. Furthermore, the errors are not symmetric; the running median of the flux ratio is consistently closer to the 84th percentile than the 16th, revealing a systematic skew where individual sources are more likely to have their flux overestimated than underestimated. Although this strategy successfully removes the large positive flux-boosting bias seen in PyBDSF (Figure \ref{fig:pybdsf_bias}), it introduces a large, asymmetric error in its place.

The performance improves significantly with the \texttt{Radio-Likely Prior}  as shown in Figures \ref{fig:sim_flux_comp_fir}. For the same 50~$\mu$Jy mask and 30 $\mu$Jy flux level, the 16th-84th percentile range narrows dramatically to just $\pm$0.15 dex. This factor of three improvement in precision demonstrates that a high-purity, albeit incomplete, prior is important for obtaining reliable individual flux measurements. We note a small but systematic negative bias of $-0.02$\,dex (a $\sim$4.5 per~cent underestimation) that is stable across the flux range, as well as a mild skew ($\sim$0.05 dex median offset) emerging at fainter flux densities. By reducing the number of faint, ambiguous priors, the algorithm converges on more stable and better-constrained solutions.

\subsubsection{The Role of the Masking Threshold}
While the prior list governs the overall precision, the choice of masking threshold affects both the precision and the accuracy of the deblended flux densities, particularly for sources near the mask limit. For both prior catalogues, masking sources $>$ 50$\mu$Jy (50$\mu$Jy mask) consistently produces better results compared to the 250$\mu$Jy mask.

When using the optimal \texttt{Radio-Likely Prior}, switching from a 50~$\mu$Jy mask (Figure~\ref{fig:sim_flux_comp_fir}, top) to a 250~$\mu$Jy mask (Figure~\ref{fig:sim_flux_comp_fir}, bottom) degrades precision. The 16th--84th percentile range at 30\,$\mu$Jy widens from 0.15\,dex to 0.25\,dex. This confirms that creating a cleaner deblending environment by removing moderately bright sources improves the precision of the remaining faint-source measurements.

The masking threshold also affects the accuracy for sources near the mask limit. When using the 250~$\mu$Jy mask, the median flux ratio decreases sharply above 60 $\mu$Jy, reaching $-0.15$ dex ($\sim$30 per~cent deficit) near the masking threshold (Fig. \ref{fig:sim_flux_comp_fir}, bottom). Extended AGNs and SFGs are particularly affected as XID+'s point-source assumption fails for their complex emission. The 50~$\mu$Jy mask eliminates this bias by excluding such sources. The 250~$\mu$Jy mask also degrades faint-source recovery. 
The 50~$\mu$Jy mask is the best choice, as it yields both higher precision for faint sources and avoids the flux underestimation bias for bright, complex sources.

\subsubsection{Masking Type: SRL vs. GAUL}
\label{sec:srl_vs_gaul}
We also tested whether the catalogue type used to generate the mask, SRL  versus the GAUL, impacts the deblended flux densities. The choice between these two is not trivial, as they are known to behave differently at both the bright and faint ends of the source population. As shown by \citet{Hale2023}, the SRL catalogue is more physically meaningful for bright, resolved radio galaxies, while the GAUL catalogue can be a better tracer of the underlying source population in the confusion-dominated faint regime.

Given these known differences, it is important to test whether the choice of masking catalogue (SRL vs. GAUL) type impacts the final deblended flux densities.
A direct comparison of the results shows that the deblending performance is qualitatively and quantitatively very similar regardless of the masking catalogue used. The 50~$\mu$Jy threshold consistently produces lower scatter and more reliable flux densities, whether it is based on the SRL or GAUL list. The reason is that the two catalogues diverge only for the very brightest, multi-component sources, which are already removed by the 50~$\mu$Jy mask. 

\subsubsection{The Optimal Strategy}
\label{sec:adopted_strategy}
The \texttt{Radio-Likely prior} provides substantially improved precision for individual flux density estimates, while the 50~$\mu$Jy mask reduces contamination from bright, complex neighbours without introducing a systematic underestimation for faint sources. Furthermore, as shown in Section \ref{sec:srl_vs_gaul}, the choice of masking catalogue (SRL vs GAUL) has a negligible impact at this threshold.  We therefore adopt the combination of the \texttt{Radio-Likely prior} with a 50~$\mu$Jy SRL-based mask for all subsequent validation.

\subsection{\textbf{Robustness Tests}}
\label{sec:robustness_tests}
Having established the optimal deblending configuration in idealized conditions, we now subject this configuration to two rigorous stress tests to evaluate its robustness against real-world observational challenges: spatially varying noise and positional uncertainties.
\subsubsection{Validation with Spatially Varying Noise}
\label{sec:sim_varying_noise}

To test the robustness of our strategy against realistic observational noise conditions, we apply our recommended strategy, the \texttt{Radio-Likely Prior} and SRL 50~$\mu$Jy mask, to a larger simulation that incorporates a realistic, spatially varying noise similar to the MIGHTEE-COSMOS field. We generated an analytic model by modelling the primary beam of each of the 22 MIGHTEE-COSMOS pointings as a Gaussian and summing their squared responses on a grid matching the real data \citep{Hale2025}. This creates an idealised weight map, which is then inverted to produce a theoretical RMS map ($\sigma_{\rm Tot} \propto 1/\sqrt{\text{weight}}$).

This process reproduces some of the features of the real observations, including the lower noise in the field centre where multiple pointings overlap, and higher noise at the edges. This pre-computed RMS map is then used as input for a \texttt{galsim} \texttt{VariableGaussianNoise} generator, which adds the appropriate level of noise to each pixel of the noiseless simulated sky \citep{Rowe2015}.

\begin{figure*}
    \centering
    \includegraphics[width=0.98\textwidth]{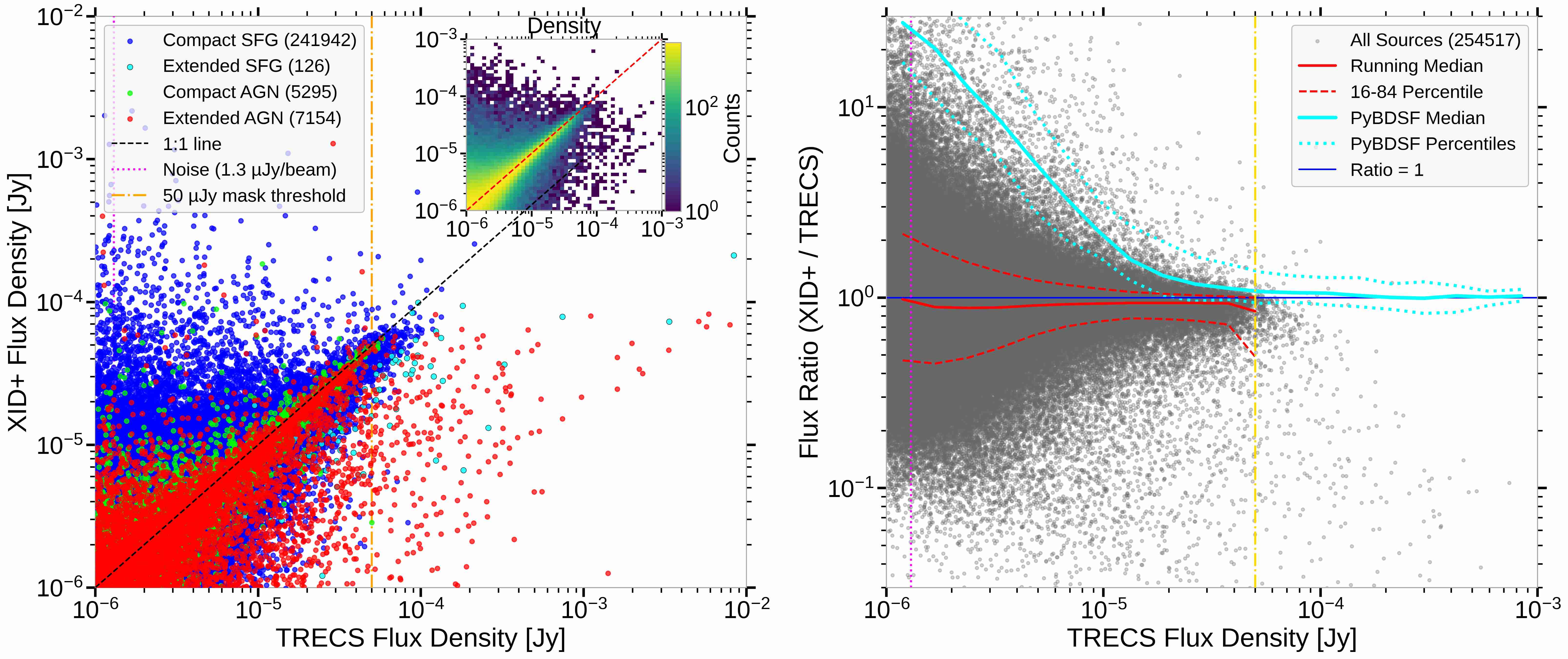}
    \caption{XID+ deblending performance on a large-scale simulation with realistic, spatially varying noise, using the \texttt{Radio-Likely Prior} prior and the 50~$\mu$Jy SRL mask. The panels and symbols are the same as in Figure \ref{fig:sim_flux_comp_full}. The flux recovery remains excellent and is highly consistent with the results from the idealized uniform-noise simulation, demonstrating the robustness of the pipeline to realistic noise conditions.}
    \label{fig:varying_noise_comp}
\end{figure*}

Figure \ref{fig:varying_noise_comp} shows the deblended flux performance for this more realistic simulation, using our recommended 50~$\mu$Jy mask. At first glance, the scatter plot for this large-scale simulation appears to show a significantly wider distribution of points than the uniform-noise case, with a greater number of outliers. However, a quantitative analysis of the bulk population reveals that the core performance remains excellent. The running median of the flux ratio remains stable, with the small systematic bias shifting only slightly from -0.02\,dex to -0.03\,dex. Similarly, the precision, as measured by the 16th-84th percentile range, widens only modestly from 0.30\,dex to 0.35\,dex at 30\,$\mu$Jy. 

This test shows that the statistical performance of the deblending remains robust against spatially varying sensitivity; the increased number of outliers reflects the larger survey area rather than a failure of the deblending algorithm.

\subsubsection{Positional Offsets }
\label{sec:positional_offsets}

In addition to noise variations, we tested the impact of positional offsets between radio and optical/infrared hosts, which can arise from both astrometric uncertainties and genuine physical displacements such as extended jets. To assess the robustness of our strategy to such offsets, we performed a dedicated stress test in which we shifted the positions of all simulated sources using an empirical offset distribution measured from the MIGHTEE Early Science multi-wavelength catalogue \citep{Whittam2024}. These offsets were derived from 6,102 sources securely matched using the likelihood ratio method, ensuring they represent the true astrometric scatter of the survey. We applied these offsets to the input catalogue before re-running our adopted configuration (\texttt{Radio-Likely} prior, 50~$\mu$Jy mask). This process reduced the number of successfully matched sources (within 2.5 arcsec) from 67,307 to 62,878, a loss of $\sim$7 per~cent, providing a realistic estimate of the catalogue yield under imperfect positional assumptions.

\begin{figure*}
    \centering
    \includegraphics[width=0.98\textwidth]{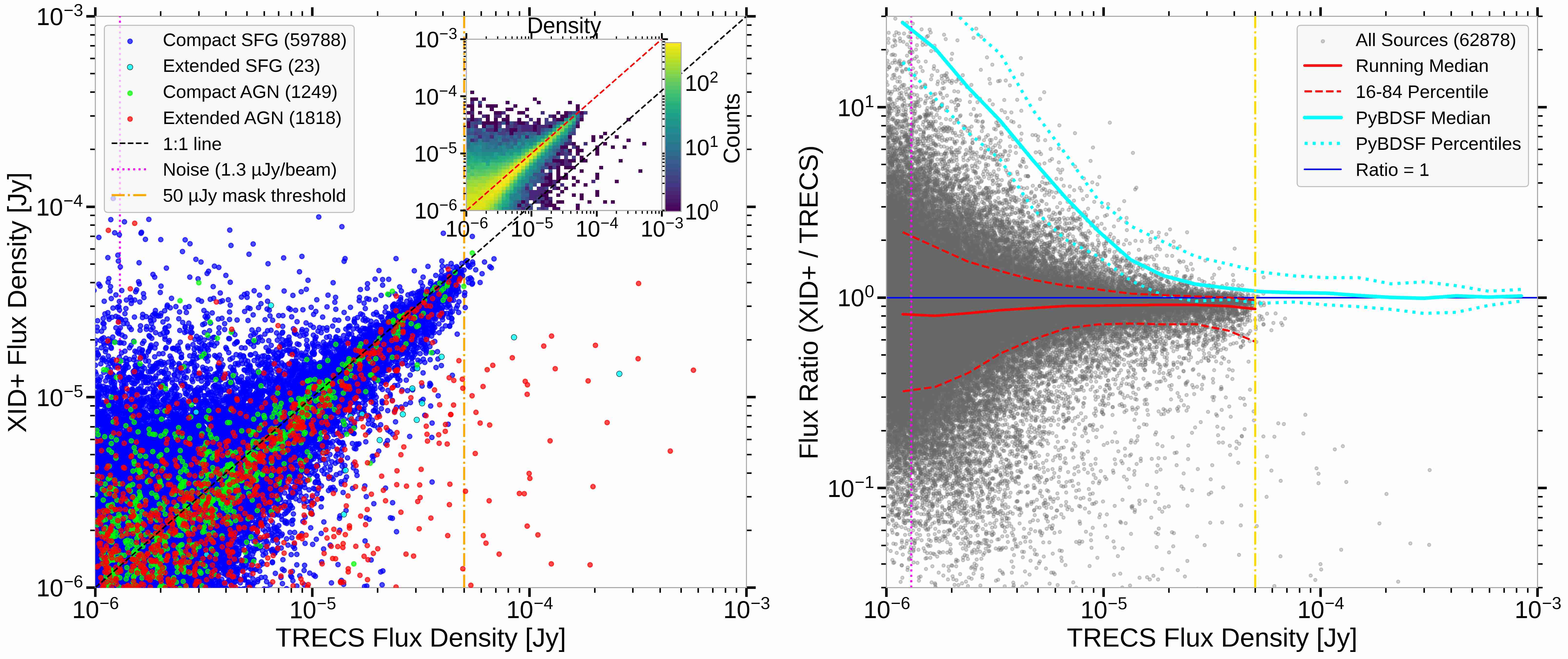}
    \caption{XID+ deblending performance using the \texttt{Radio-Likely Prior} and a 50~$\mu$Jy SRL mask, where the input prior positions were shifted using empirical offsets drawn from the MIGHTEE Early Science multiwavelength catalogue. The panels and symbols are the same as in Figure \ref{fig:sim_flux_comp_full}. \rev{This test demonstrates the pipeline's robustness against positional offsets between radio and optical/infrared positions, which can arise from astrometric uncertainties or genuine physical displacements.}
 }
    \label{fig:pos_offset_comp}
\end{figure*}

As shown in Figure \ref{fig:pos_offset_comp}, at faint flux densities ($S \lesssim 4\,\mu$Jy), the recovered flux density is the most sensitive to positional uncertainties. At \rev{$S \sim1.8\,\mu$Jy}, the median recovered-to-true flux ratio decreases from 0.90 in the ideal simulation to 0.80 when offsets are applied, doubling the bias. The scatter similarly increases from 0.54 dex to 0.73 dex. 
When comparing this to other configurations/tests, the positional offsets produce a larger scatter and bias than either the 250~$\mu$Jy mask (0.64 dex scatter) or the spatially varying noise simulation (0.53 dex scatter). However, the performance remains superior to the \texttt{Full Prior}, which suffers from a scatter of $\sim0.90$ dex in this regime.

At intermediate flux densities ($S \approx 10\,\mu$Jy), the effect of positional offsets is substantially weaker. The median ratio changes only slightly from 0.94 to 0.91, and the scatter remains tight, increasing from 0.15 dex to 0.18 dex. In this regime, the choice of prior remains the dominant factor governing accuracy. The simulation with positional offsets yields a precision comparable to both the 250~$\mu$Jy mask and the spatially varying noise runs ($\sim0.16$ to $0.19$ dex). For bright sources ($S \gtrsim 30\,\mu$Jy), the performance is nearly unchanged.

In summary, while positional offsets represent the main source of scatter at the faintest flux densities, their relative impact decreases rapidly with increasing S/N. Although applying these shifts yields higher scatter than the idealized simulations, the performance remains better than that obtained using the full T-RECS prior. This demonstrates that the adopted \texttt{Radio-Likely Prior} with a 50~$\mu$Jy mask is robust to the level of positional mismatch expected in real observations, ensuring reliable flux recovery for the bulk of the source population.

\subsection{The Reliability of Individual Deblended Flux Densities}
\label{sec:sim_reliability}

The stress tests in Section~\ref{sec:robustness_tests} confirm that our optimal strategy performs well under realistic noise and positional offsets. We now examine the statistical reliability of the individual flux measurements themselves, focusing on uncertainty calibration, posterior asymmetry, detection significance, and goodness-of-fit.

\subsubsection{Flux Density and Uncertainty Recovery}
\label{sec:stat_validation} 
\begin{figure}
    \centering
    \includegraphics[width=0.48\textwidth]{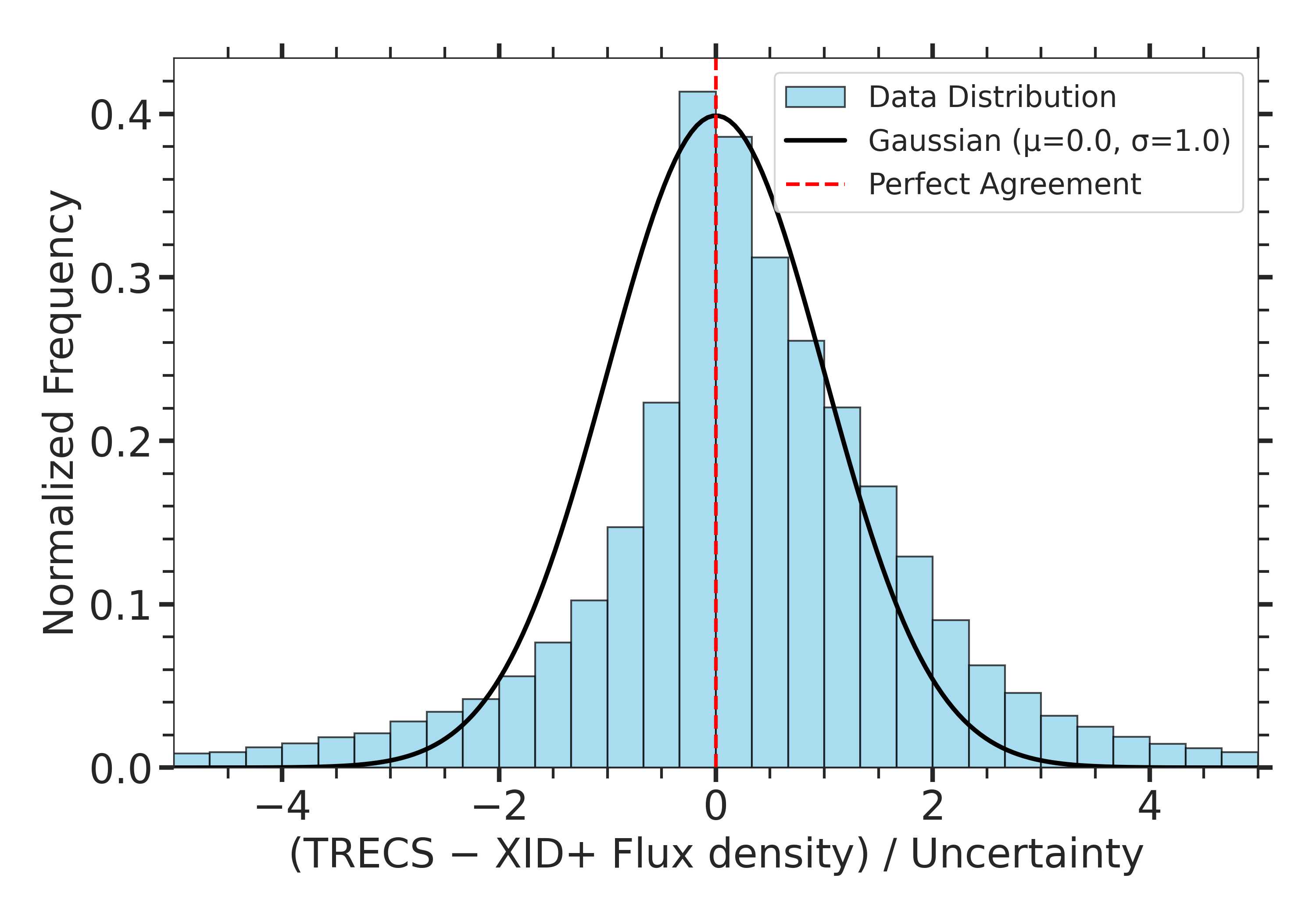}
    \caption{The normalized residual distribution for the simulated sources. The x-axis shows the difference between the true T-RECS flux and the recovered XID+ flux, divided by the XID+ uncertainty. The blue histogram shows the distribution for our data. The solid black line is a standard normal (Gaussian) distribution with a mean ($\mu$) of 0.0 and a standard deviation ($\sigma$) of 1.0, which represents the ideal result. The red dashed line indicates the ideal mean of zero.}
    \label{fig:flux_validation}
\end{figure}

An informative test of the deblending framework is to assess whether the recovered flux densities are unbiased relative to their estimated uncertainties. We construct a normalized residual distribution, shown in Figure~\ref{fig:flux_validation}, which plots the quantity (True Flux density - XID+ flux density) / (XID+ Uncertainty). Because XID+ reports asymmetric posterior intervals, we adopt a conservative, single-value uncertainty per source defined as the larger of the upper (84th - 50th percentile) and lower half-intervals (50th - 16th percentile). Using the larger half-interval ensures that the significance is not overstated for skewed posteriors and makes the test conservative. For unbiased measurements with correctly calibrated uncertainties, this distribution should follow a Gaussian with mean zero and unit variance.

The observed distribution is complex. The peak is shifted slightly to the left of zero, while it also shows an excess of sources with positive residuals compared to the ideal Gaussian, particularly in the 0 to +2 range. These competing features result in a global mean for the distribution that is extremely close to zero. This demonstrates that while there are competing statistical effects on individual sources, the catalogue as a whole is unbiased.

This finding is consistent with the flux ratio analysis in Section \ref{sec:prior_impact}, which used the running median and found a small systematic underestimation of $\sim$4.5 per~cent. The fact that different statistical estimators (the mean of the residuals versus the median of the flux ratio) on this non-Gaussian distribution both show that any systematic bias is minor, at the few-percent level, provides strong support for the reliability of the results.

\subsubsection{Individual Flux Density Uncertainties}
\begin{figure}
    \centering
    \includegraphics[width=0.9\linewidth]{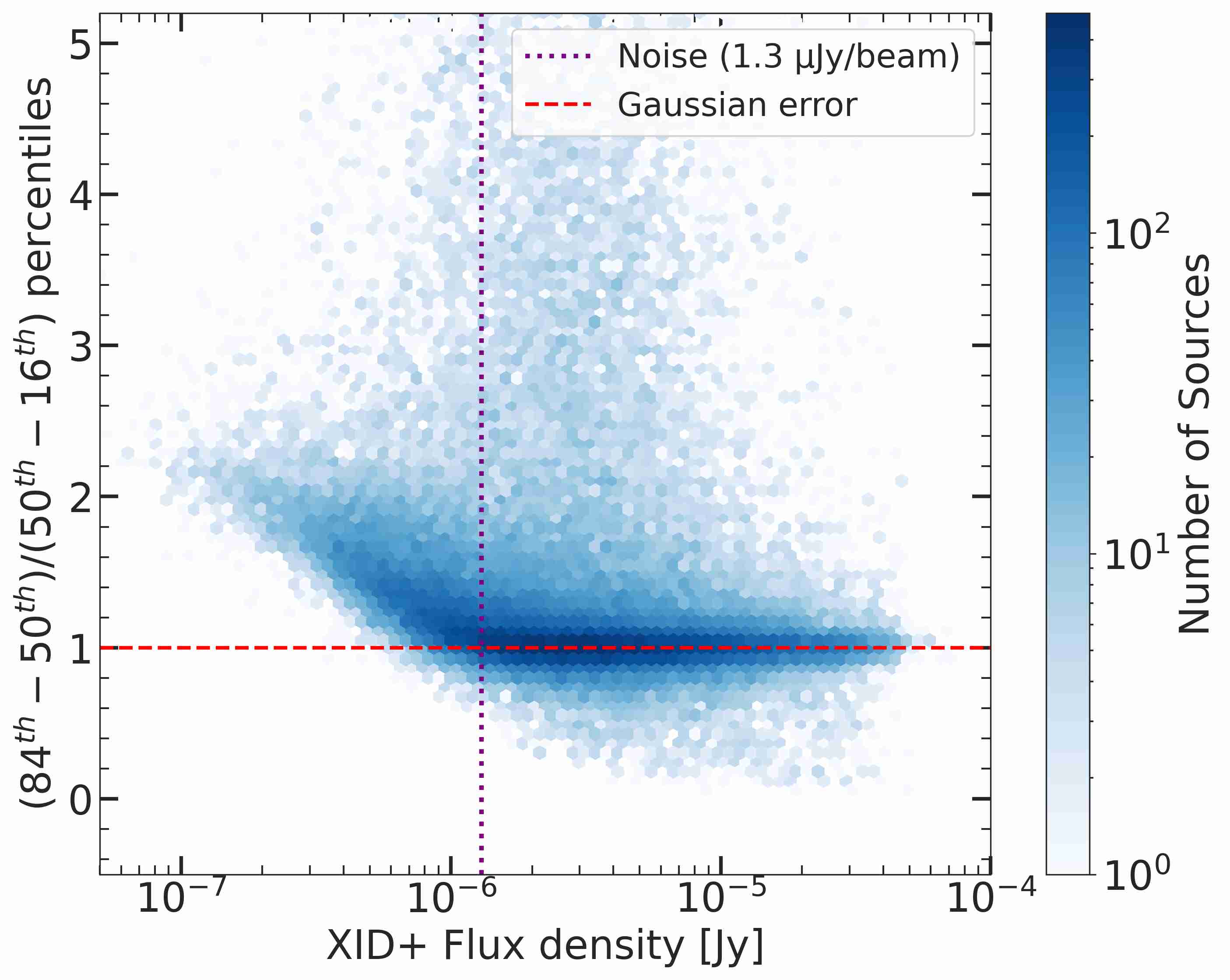}
    \caption{ The asymmetry of the posterior uncertainty as a function of deblended flux density for the 50~$\mu$Jy mask and \texttt{Radio-Likely Prior}. The y-axis shows the ratio of the upper (84th - 50th) to the lower (50th - 16th) percentile intervals. A ratio of 1 (red dashed line) indicates symmetric (Gaussian-like) posteriors. The bulk of the source population remains symmetric down to flux densities below the noise limit (\rev{1.3}\,$\mu$Jy\rev{/beam}, purple dotted line).}
    \label{fig:uncertainty_asym}
\end{figure}

To understand the nature of the individual uncertainties, we examine the asymmetry of the posterior probability distributions, as shown in Figure \ref{fig:uncertainty_asym}. We quantify the shape of the posterior by the ratio of the upper uncertainty (84th - 50th percentile) to the lower uncertainty (50th - 16th percentile). A ratio of 1 indicates a symmetric, Gaussian-like posterior. For the vast majority of sources, the posterior remains symmetric (ratio $\approx$ 1) down to flux densities comparable to, and even slightly below, the \rev{1.3}\,$\mu$Jy\rev{/beam} (1$\sigma$) noise limit. This indicates that for most sources at the noise level, the uncertainties are well-behaved. For faint sources below the noise limit, the ratio becomes $\gg 1$. This is an expected consequence of the posterior being truncated by the physical non-negativity constraint on flux density, resulting in a skewed distribution with a long tail towards higher values. Interestingly, the plot also reveals a population of sources at intermediate flux densities that exhibit highly skewed uncertainties (ratio > 2). \rev{This behaviour is expected: at lower flux densities, the posterior becomes increasingly affected by the physical non-negativity constraint on flux density, leading to asymmetric uncertainties. This is a natural consequence of lower S/N rather than an indication of unreliable measurements.}

\subsubsection{Assessing Detection Significance}
\label{sec:detection_significance}

\begin{figure}
    \centering
    \includegraphics[width=\columnwidth]{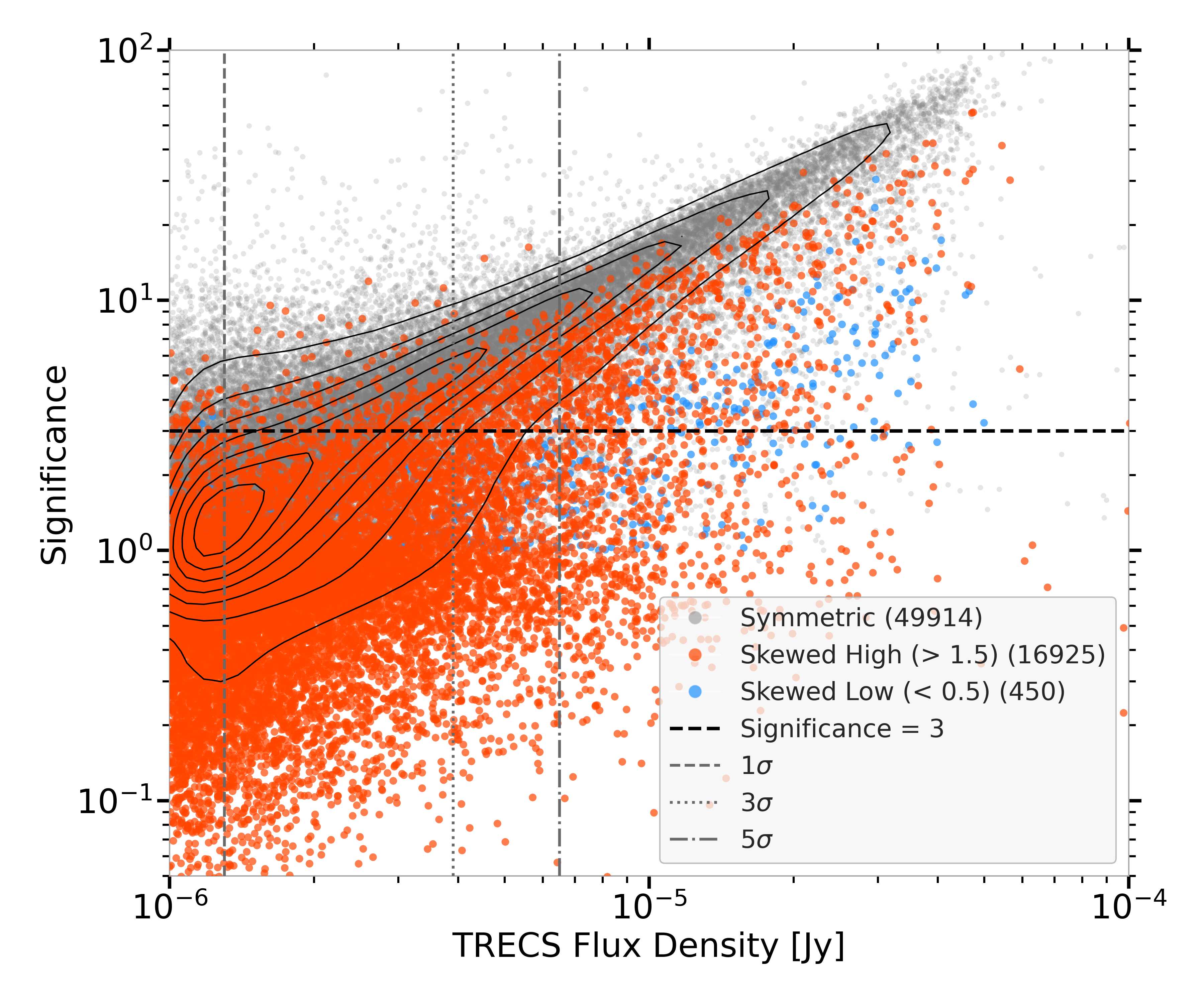}
    \caption{Recovered significance (median flux density divided by conservative uncertainty) as a function of true T-RECS input flux density for the \texttt{Radio-Likely Prior}, and 50~$\mu$Jy SRL mask. Points are colour-coded by posterior asymmetry: grey = symmetric posteriors, orange = skewed to high flux, blue = skewed to low flux, with black contours indicating the density of the distribution. The vertical dashed lines indicate the 1$\sigma$, 3$\sigma$, and 5$\sigma$ thermal noise levels of the simulation, while the horizontal red dashed line indicates a significance of 3. 
}
    \label{fig:significance_plot}
\end{figure}

The presence of sources with skewed or asymmetric posteriors at all flux density levels, as identified in the previous section, motivates an assessment of how posterior skewness affects detection reliability. We investigate this by examining the recovered statistical significance of each source against its true input flux density in Figure \ref{fig:significance_plot}. We define the significance as the ratio of the median recovered flux density to its conservative uncertainty (defined in Section \ref{sec:stat_validation} as the maximum of the upper and lower error bounds). This ensures that we do not overestimate the significance of sources with skewed posterior distributions.

The result is a positive correlation between the true flux density and the recovered significance across the entire flux density range. The sources with skewed posteriors (orange and blue points) follow the same statistical trend as the symmetric ones (grey points); but cover a wider range of significance, which increases with decreasing flux density. This shows that a simple flux density cut is not a sufficient criterion for selecting a sample of well-constrained sources. For example, at a true flux density of 3.9\,$\mu$Jy (the 3$\sigma$ limit), sources can have significances ranging from less than 1 to over 10, regardless of their posterior shape.

Based on this, we provide two recommendations for using the deblended catalogue. For studies that can incorporate the full posterior probability distribution for each source, the entire catalogue can be used, as the posteriors correctly capture the complex uncertainties. However, for many science cases, a simple, high-purity sample is required. For these applications, we recommend a selection based on significance. A cut of Significance $>$ 3 provides a practical way to flag a reliable subset of detections. Given its usefulness for assessing detection reliability, we will provide the significance for each source in the final deblended catalogue.

\subsubsection{Goodness-of-Fit: The p-value Residual Statistic}
To assess the goodness-of-fit for each source, we use the Bayesian p-value residual \citep{Gelman1996} statistic, as defined in \citet{Shirley2021}. This metric is a form of posterior predictive check that quantifies model fit to local data given the uncertainties. The resulting statistic is a probability, where a value of 0 indicates a model that always provides a good fit, and a value of 1 indicates a model that always provides a poor fit.

\begin{figure*}
    \centering
    \includegraphics[width=0.49\textwidth]{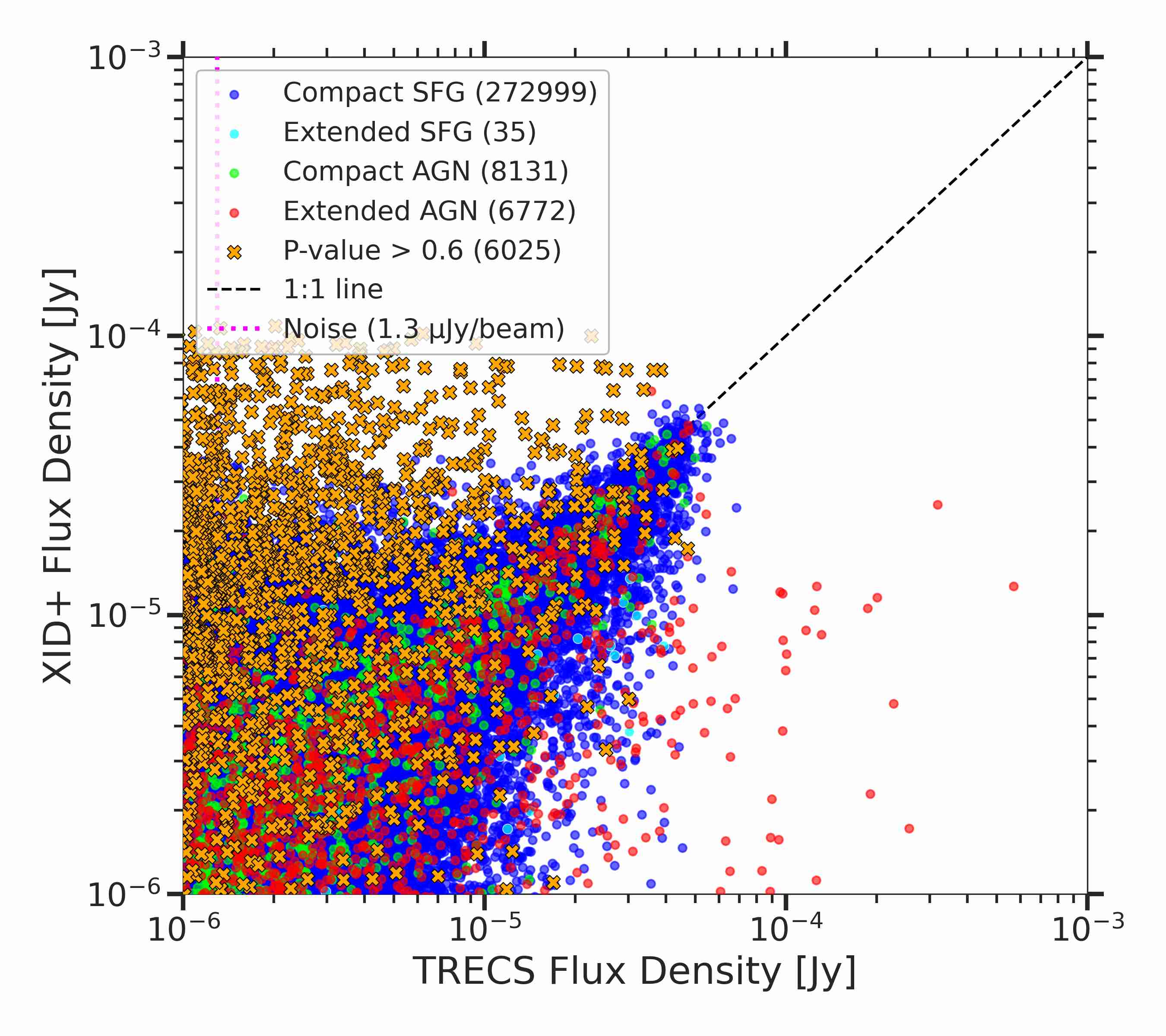}
    \includegraphics[width=0.49\textwidth]{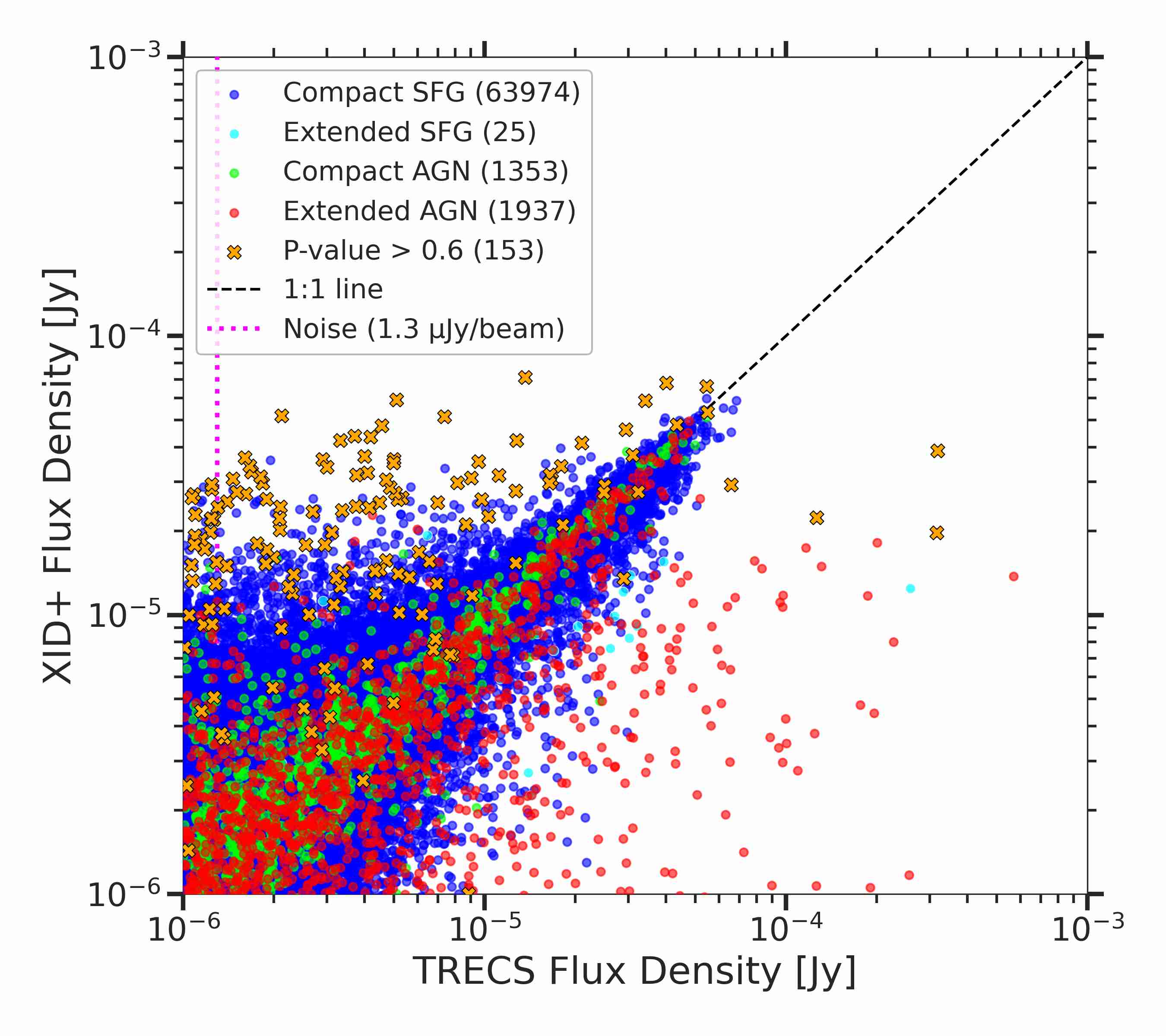}
    \caption{The effect of the Bayesian p-value as a quality flag, using the 50~$\mu$Jy SRL mask. Sources with unreliable fits (p > 0.6) are highlighted in orange. {Left:} Using the full T-RECS prior, a number of sources are flagged. \textit{Right:} Using the \texttt{Radio-Likely Prior} , the fraction of unreliable fits is almost negligible.  This shows that a high‑purity prior leads to more robust fits and that the p‑value statistic is an effective diagnostic for identifying outliers.  
    }
    \label{fig:pvalue_cut}
\end{figure*}

We identify sources with a p-value residual statistic > 0.6 as having an unreliable fit. This threshold is determined empirically from our simulations (see Figure \ref{fig:pvalue_cut}), where it was found to effectively flag sources with significantly biased flux densities without removing a large fraction of well-fit sources. High p-values may arise when sources are poorly described by XID+'s point-source assumption, such as bright, extended sources that violate this model.

The power of the recommended strategy, \texttt{Radio-Likely Prior}  and a 50~$\mu$Jy mask, is evident in the right panel as the fraction of sources flagged with unreliable p‑values is reduced. This shows that the strategy improves both the overall flux accuracy and precision and the reliability of individual fits.

\subsection{Recovery of Source Counts}
\begin{figure*}
    \centering
    \includegraphics[width=1.95\columnwidth]{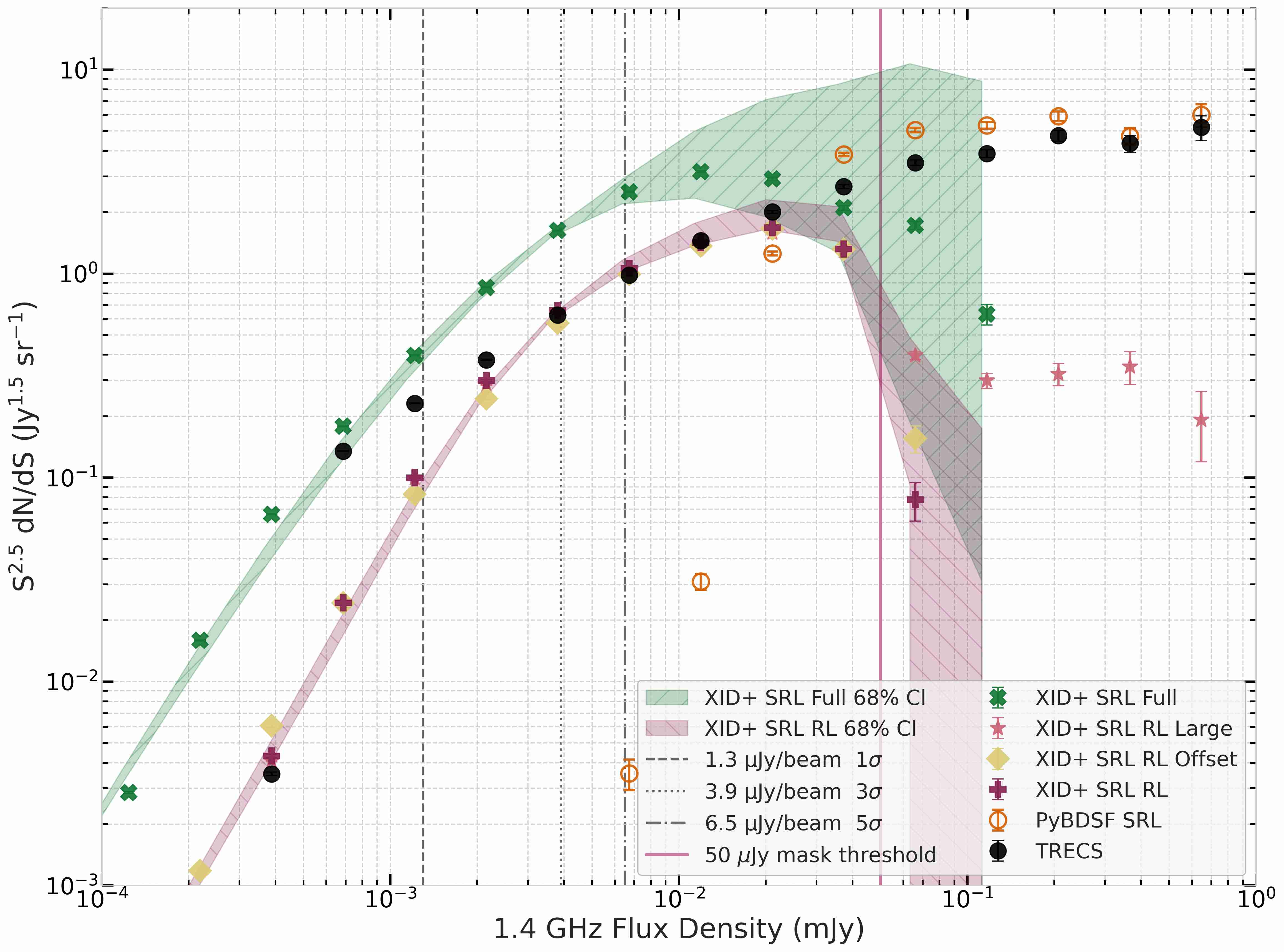}
    \caption{Euclidean-normalized 1.4 GHz source counts from the simulated data, comparing various extraction methods to the ground truth. The ground truth, injected source counts from the T-RECS simulation, are shown as solid black circles. The source counts extracted from our simulated map using the PyBDSF SRL catalogue are shown as unfilled orange circles. The deblended results for the \texttt{Full Prior} are shown as green crosses. For the \texttt{Radio-Likely Prior}, results are shown as dark red plus-signs. The deblended counts from our large-area, varying-noise simulation are shown as red stars. The results from the simulation incorporating positional offsets are shown as peach diamonds. All XID+ extraction runs shown here use the 50~$\mu$Jy mask. The shaded green and pink regions represent the 68 per~cent confidence intervals for the \textit{Full Prior} and \texttt{Radio-Likely Prior} deblending runs, respectively. The vertical grey lines indicate 1, 3, and 5 times the thermal noise ($\sigma$ = 1.3 $\mu$Jy\rev{/beam}). The vertical red line indicates the upper flux limit of 50~$\mu$Jy; sources brighter than this were masked prior to deblending.}
    \label{fig:source_counts_sim}
\end{figure*}

A robust test of the entire process is to compare the Euclidean-normalized source counts ($S^{2.5} dN/dS$) of the recovered catalogue against the true input catalogue to the simulation, where the injected source population is limited to flux densities above 0.5~$\mu$Jy. Figure \ref{fig:source_counts_sim} shows the source counts. First, we examine the results from our PyBDSF blind source extraction. At bright flux densities ($S_{1.4} \gtrsim 0.5$ mJy), the SRL counts are in excellent agreement with the true T-RECS counts. 

At intermediate flux densities ($S_{1.4} \lesssim 0.5$ mJy), the PyBDSF SRL catalogue shows a significant upturn, a clear signature of flux boosting. Below $\sim 0.2$~mJy, the PyBDSF counts drop off sharply as the sources in this flux range are boosted to higher flux densities.

Next, we analyze the deblended counts, which are defined only below their respective masking thresholds. The results from using the \texttt{Full Prior} show a significant overestimation of the true T-RECS counts. This is because when presented with a dense, complete list of all sources, the algorithm fits low-level flux to the numerous priors, essentially promoting noise peaks into spurious low-flux detections. This effect is more pronounced for the less restrictive 250~$\mu$Jy mask, which provides a more complex environment for the algorithm.

The counts from our recommended configuration, using the \texttt{Radio-Likely Prior}, accurately trace the true T-RECS distribution. For the 50~$\mu$Jy mask, the deblended counts are in excellent agreement with the ground truth from ~30 $\mu$Jy down to approximately 3.9 $\mu$Jy (3$\sigma$). For the  250~$\mu$Jy mask, the counts are reliable from approximately 100 $\mu$Jy, to 3 $\sigma$. In both cases, the counts drop off sharply below ~2 $\mu$Jy, due to the intrinsic incompleteness of the prior itself, which was defined with a 1 $\mu$Jy flux cut.

To quantify the statistical reliability of these results, we derive their uncertainties directly from the full posterior probability distributions. As the deblending was performed on independent tiles, the MCMC chains for each tile are statistically independent. To propagate this uncertainty, we construct a global confidence region by ranking each of the 1000 MCMC samples by their log-posterior probability within each tile. The 68 per~cent confidence interval is then defined by the envelope between the source count curves derived from the highest-ranked ("best-fit") and lowest-ranked ("worst-fit") global samples. To avoid over-cluttering the plot, we show only these confidence intervals for the primary 50~$\mu$Jy deblending runs. 
As shown in Figure \ref{fig:source_counts_sim}, the median-derived source counts for both the \textit{Full Prior} and \texttt{Radio-Likely Prior} lie well within their respective 68 per~cent confidence intervals. A notable feature is that the confidence intervals widen as they approach the 50~$\mu$Jy masking threshold. 

Finally, results from both the large-area, varying-noise simulation (red stars) and the positional offset test (peach-yellow diamonds) show excellent agreement with the ideal case, demonstrating that our statistical determination of source counts is robust to realistic observational uncertainties.

\subsection{Summary and Recommended Strategy}
\label{sec:summary}
The simulations in Section~\ref{sec:sim_framework} provide clear guidance for applying XID+ to the MIGHTEE survey. They show that deblending is essential, as standard source-finding with PyBDSF suffers from significant flux boosting in the confused regime, making it unsuitable for accurate faint-source science. The quality of the prior list is the dominant factor: a high‑purity, albeit incomplete, \texttt{Radio-Likely Prior}, yields far more accurate flux density measurements, with a scatter of 0.3\,dex at 30\,$\mu$Jy, than a complete but unfiltered prior, which has a scatter greater than 1\,dex. These tests show that a 50~$\mu$Jy masking strategy is the best, as it improves faint-source precision and avoids a systematic underestimation bias for bright, extended sources. When combined, this optimal strategy accurately recovers the true source number counts down to the $\sim 3\sigma=$3.9\,$\mu$Jy.

This performance remains stable under realistic observational conditions; our stress tests with spatially varying noise and empirical positional offsets confirm that the strategy is robust to these real-world imperfections. Finally, the Bayesian p-value provides a necessary diagnostic for identifying the small fraction of poorly constrained sources. Based on these findings, our recommended strategy for the real MIGHTEE data is to use a high-confidence, \texttt{Radio-Likely Prior} list and apply a mask to exclude sources $>$ 50~$\mu$Jy. We further recommend that the final catalogue include the Bayesian p-value and detection significance to facilitate the selection of reliable samples.

\section{Application to the MIGHTEE-COSMOS Field}
\label{sec:application}

Having established and tested our optimal deblending strategy on realistic simulations, including spatially varying noise and positional offsets, we now apply this approach to the MIGHTEE‑COSMOS low‑resolution map. This section details the construction of the final prior catalogue and mask, and presents the results of the deblending; the outcome is a high-fidelity source catalogue and number counts.

\subsection{Prior List}

Based on the simulation results in Section \ref{sec:sim_validation}, we adopt the \texttt{Radio-Likely Prior} as part of our deblending strategy for the primary scientific analysis. The results from this high-purity approach are presented in the main body of this section. For completeness, the results from the \textit{Comprehensive L\textsubscript{IR} Prior} are presented in Appendix \ref{appendix:lir}, where they serve as a valuable real-world confirmation of our simulation findings.

\subsection{Masking Strategy}
We applied a multi-layered masking strategy to the real data to exclude regions unsuitable for analysis. This includes masking for radio artefacts and regions of incomplete multi-wavelength data.

\subsubsection{Radio Masking}
Following the results from our simulations, our primary mask is generated from the PyBDSF SRL catalogue by masking all sources brighter than 50~$\mu$Jy. Specifically, we select all sources from the MIGHTEE DR1 SRL catalogue from \cite{Hale2025} with a peak flux greater than 50~$\mu$Jy and mask them. 

Beyond the primary mask validated in simulations, we adopt a more aggressive elliptical mask around the very brightest sources ($S_{1.3, \rm peak} > 6$\,mJy). This is necessary to address observational artefacts, such as residual sidelobes from imperfect deconvolution, which are not present in our idealized simulations. These artefacts can extend over larger areas than the source emission itself. We determined an empirical relationship between source flux and the required elliptical mask size by measuring the extent of these artefacts around several bright sources in the MIGHTEE-COSMOS map. This relationship defines progressively larger mask radii for sources in different flux density bins (Table \ref{tab:flux_mask}). The final radio mask is the combination of the 50~$\mu$Jy model-based mask and this additional artefact mask.

\begin{table}
\centering
\caption{Flux-dependent elliptical mask radii and the number of sources in each flux range (\textbf{N}). Radius in arcmin.}
\label{tab:flux_mask}
\begin{tabular}{cccc}
\toprule
\textbf{Flux Range (mJy)} & Minor Radius & Major Radius & \textbf{N} \\
\midrule
6.0 $\le$ $S_{1.3 \mathrm{ GHz}}$ < 12.0 & 0.81 & 1.09 & 48 \\
12.0 $\le$ $S_{1.3 \mathrm{ GHz}}$ < 19.0 & 0.98 & 1.77 & 16 \\
19.0 $\le$ $S_{1.3 \mathrm{GHz}}$ < 25.0  & 1.50 & 2.47 & 5  \\
$S_{1.3 \mathrm{GHz}}$ $\ge$ 25.0        & 1.53 & 2.97 & 20 \\
\bottomrule
\end{tabular}
\end{table}

\subsubsection{Stellar Masking}
Finally, we mask regions around bright stars using the COSMOS2020 combined star mask \citep{Weaver2022}. We mask these regions because the multi-wavelength data used to define our prior are incomplete in these areas.

\subsubsection{Final priors}
Applying a combined mask, which incorporates both our radio masks and the star masks \citep{Weaver2022}, to our two raw prior lists yields the final input catalogues used for deblending.  The \textit{Comprehensive L\textsubscript{IR} Prior} is reduced to 298,894 sources. The more targeted \texttt{Radio-Likely Prior}  is reduced to 89,562 sources. 

\subsection{Flux Recovery and Comparison}
\label{sec:real_flux_recovery}
We ran XID+ on the MIGHTEE-COSMOS map using the \texttt{Radio-Likely Prior}, resulting in a full deblended catalogue of 89,562 sources. The primary diagnostic for the reliability of individual flux measurements is the Bayesian p-value residual statistic. Our initial deblending run, using sampler settings identical to those validated in our simulations (i.e., the \texttt{Stan} default \texttt{max\_treedepth=10}), led to a large number of sources with poor goodness-of-fit statistics (9,238 sources, or over 10 per cent of the sample, had a Bayesian p-value > 0.6). 

Further investigation showed that un-modeled artefacts and extended emission in the real MIGHTEE data produced complex posteriors that the sampler could not explore fully at the default depth, resulting in incomplete MCMC convergence. Increasing the \texttt{max\_treedepth} parameter to 15 allowed the NUTS sampler to explore these posteriors more thoroughly. This computationally expensive change dramatically improved the fits, reducing the number of poorly fit sources by over 68 per cent (from 9,238 to 2,933), now representing just 3.3 per cent of the sample.

To validate our flux densities, we compare them to the independent analysis of the MIGHTEE data using the super-deblended technique \citep{Jin2018, Liu2018}. \rev{This approach performs PSF-fitting at the (super-deblended) prior positions described in Section \ref{sec:prior_validation}. The resulting catalogue \citep{Sillassen2024} provides flux densities and errors that were rigorously calibrated to quasi-Gaussian uncertainties using Monte Carlo simulations performed directly on the real map \citep{Jin2018, Liu2018}. This catalogue has been widely used in many works \citep[e.g.,][]{An2021, Delvecchio2021, Jin2022,Jin2024, Sillassen2024}.}

\begin{figure}
\centering
\includegraphics[width=0.9\columnwidth]{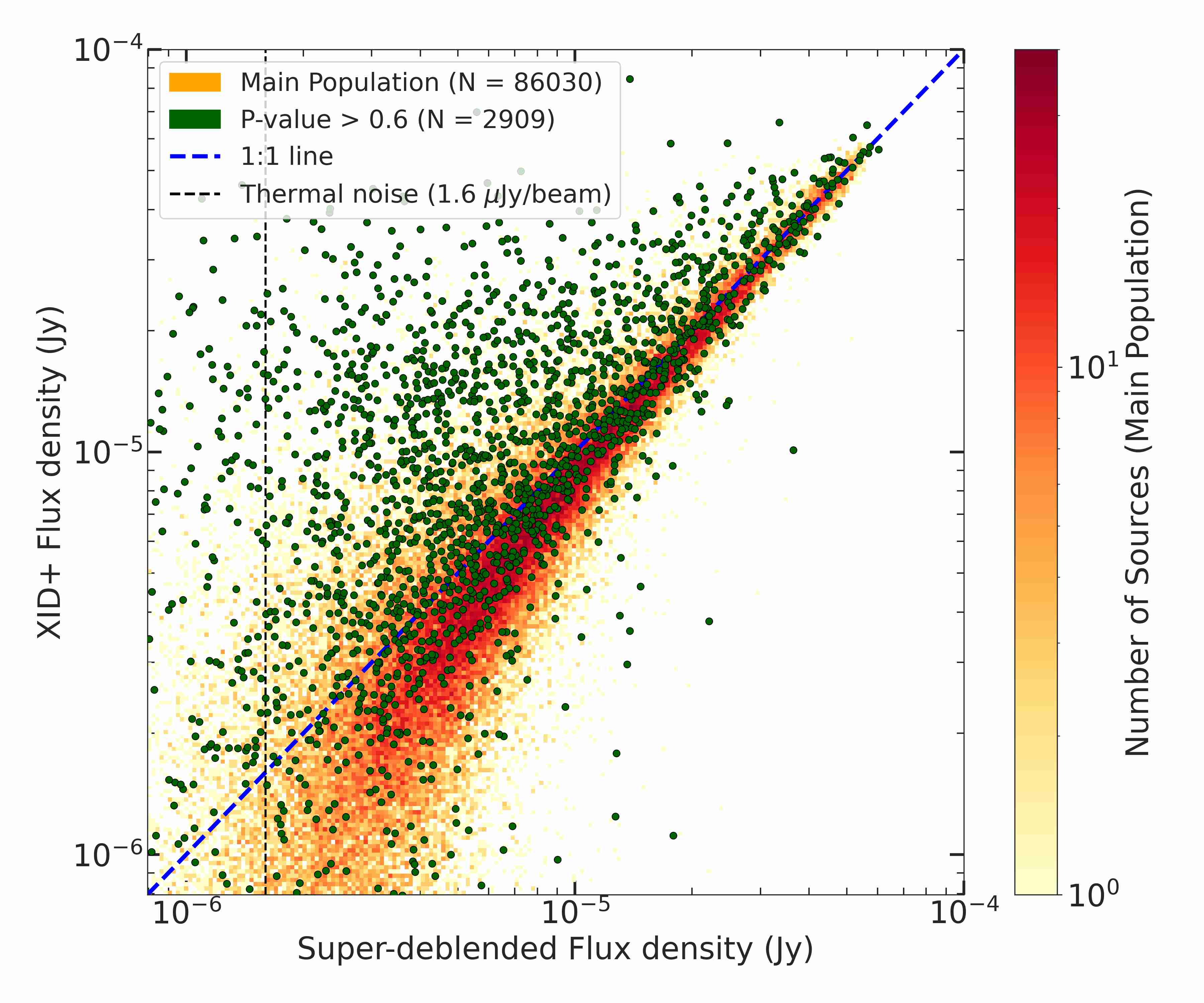}
\caption{Comparison between XID+ \texttt{Radio-Likely Prior} flux densities from this work and the super-deblended flux densities \citep{An2021,Sillassen2024} for 88,939 matched sources. Sources with an unreliable fit (Bayesian p-value > 0.6) are highlighted as dark green markers.}
\label{fig:flux_comp_fir_highlighted}
\end{figure}

We cross-matched our full \textit{Radio-Likely Prior} catalogue (89,562 sources) with the super-deblended catalogue using a 0.3 arcsec radius, finding 88,939 matches. As shown in Figure \ref{fig:flux_comp_fir_highlighted}, the two methods show excellent agreement for the bulk of the population, as they lie tightly along the 1:1 line with some scatter. The sources flagged with a high p-value in our analysis (dark green points) are mostly those that form the outlier population where our XID+ flux densities are systematically higher than the super-deblended values. This confirms that the p-value is an effective diagnostic for identifying potentially unreliable flux measurements.

Our final data product is a complete catalogue of 89,562 sources. For each source, we provide the Bayesian p-value statistic, which assesses the goodness of fit, and the detection significance. We do not remove sources from the catalogue, as the full posterior distributions for low-significance sources remain statistically valid for stacking or ensemble analyses.

However, for applications requiring robust individual source detections, we first consider a simple significance‑based selection. Applying the criterion defined in Section \ref{sec:detection_significance} (Significance $>3$) yields 31,779 sources, while a flux density cut above the average $3\sigma$ limit ($4.8\,\mu$Jy) yields 31,083 sources. To construct a highly conservative recommended sample, we select sources that satisfy both of these criteria simultaneously and exhibit a reliable goodness-of-fit (p-value residual statistic $< 0.6$). This rigorous selection results in a high-fidelity catalogue of 20,757 sources.

\subsection{Uncertainty Characterization}
\begin{figure}
\centering
\includegraphics[width=0.8\columnwidth]{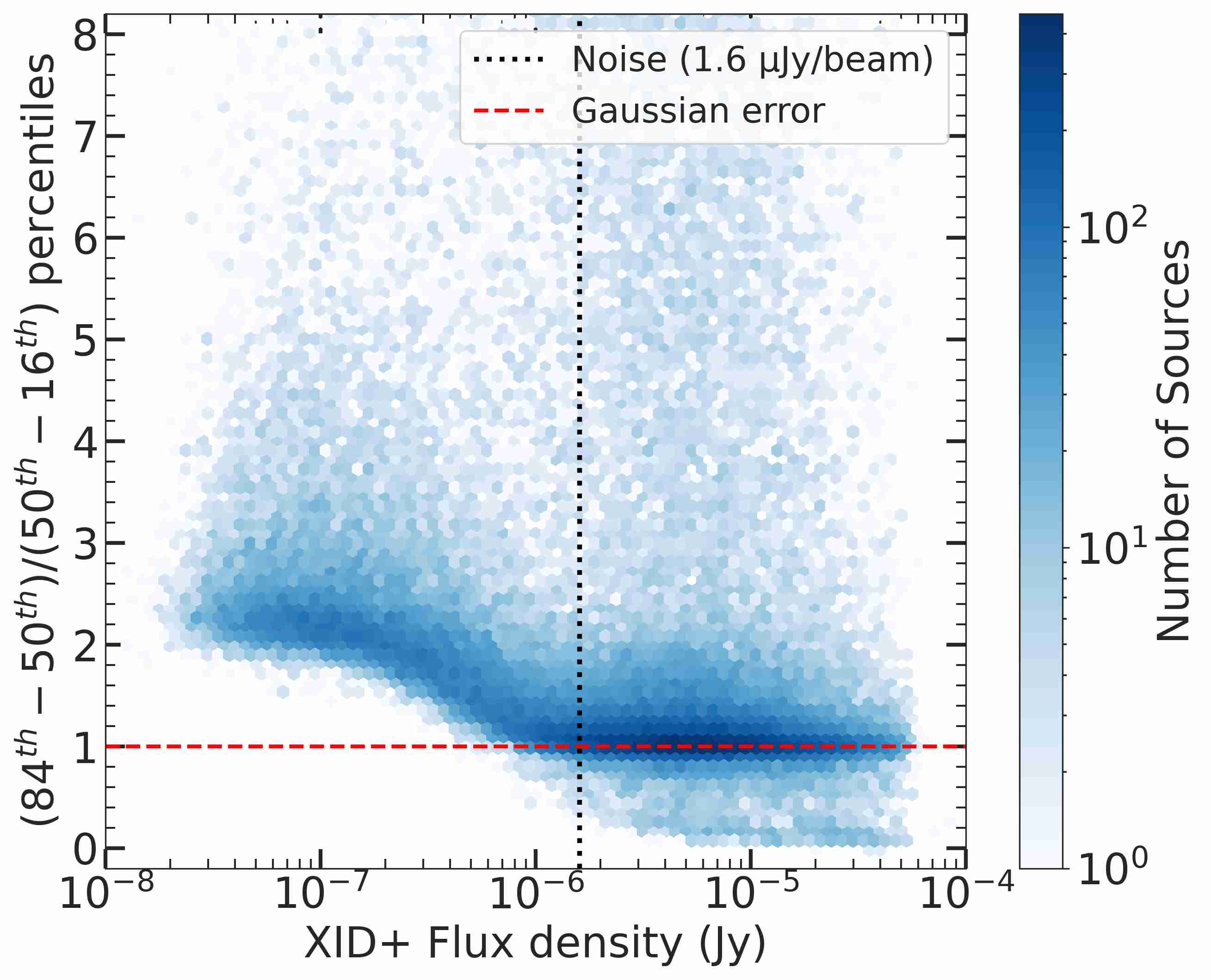}
\caption{The ratio of the upper to lower flux density uncertainty as a function of the deblended XID+ flux for sources from the \texttt{Radio-Likely Prior}. The ratio increases for faint sources, correctly capturing the skewed posterior distribution near the noise limit.}
\label{fig:uncertainty_FIR}
\end{figure}

As predicted by our simulations (Section \ref{sec:sim_reliability}), the flux density uncertainties for the real data exhibit a skewed nature for faint sources near the noise limit, reflecting the physical non-negativity constraint on flux density. Figure \ref{fig:uncertainty_FIR} demonstrates this behavior, showing that the ratio of the upper-to-lower uncertainty is close to 1 for bright sources but increases systematically as flux densities approach the thermal noise ($\sigma$) of the map. This confirms that our analysis captures the expected complex posterior distributions for the faint source population.

\subsection{Radio Source Number Counts}
\begin{figure*}
\centering
\includegraphics[width=1.95\columnwidth]{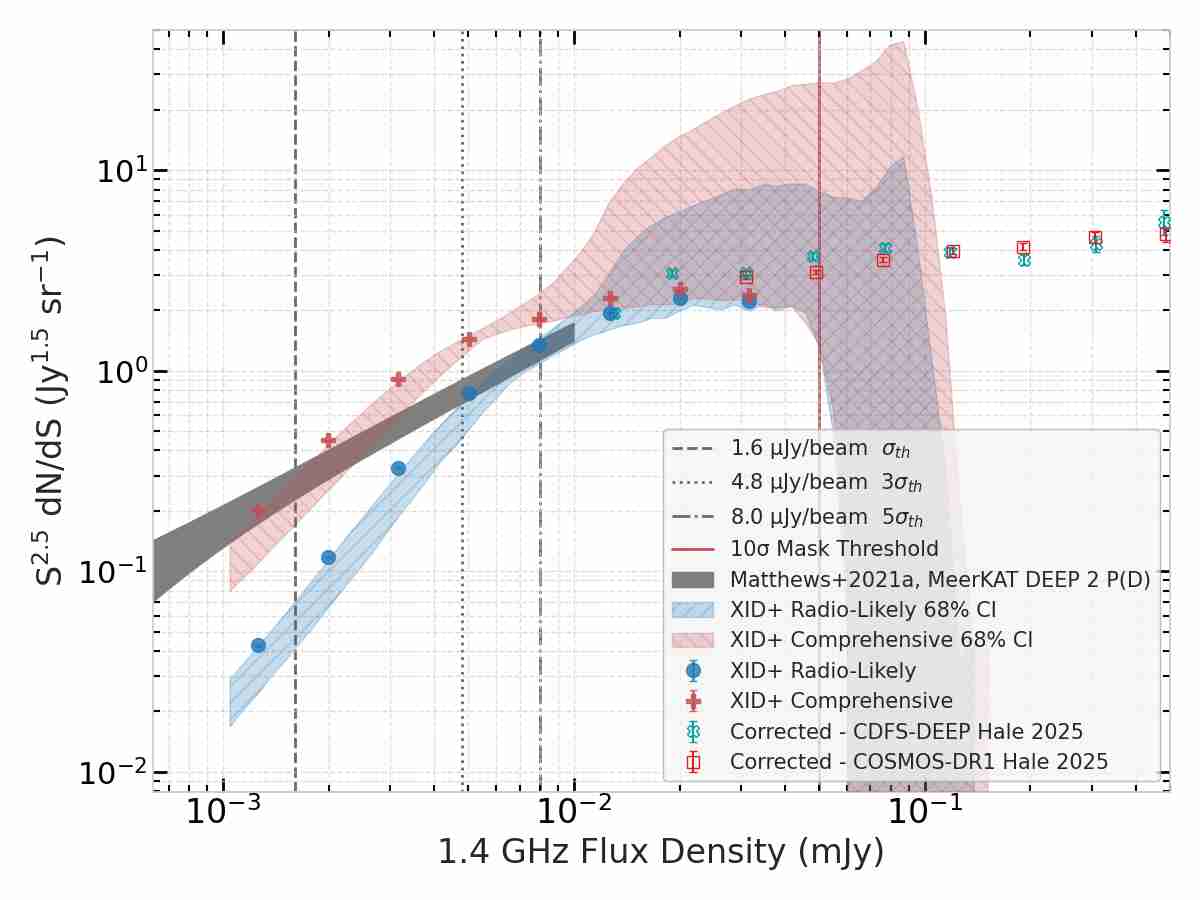}
\caption{Euclidean-normalized 1.4 GHz radio source counts. The deblended counts from our optimal \texttt{Radio-Likely Prior} are shown as blue filled circles, while the results from the \textit{Comprehensive L\textsubscript{IR} Prior} are shown as red crosses. The blue and red shaded regions represent the 68\% confidence intervals for these measurements, respectively. For comparison, we plot several data points from recent literature. From the MIGHTEE DR1 \citep{Hale2025}, the completeness-corrected counts for the COSMOS (red unfilled squares) and the CDFS-DEEP (cyan x-markers). The grey shaded region represents the P(D) analysis of the MeerKAT DEEP2 field from \protect\cite{Matthews2021}. Vertical dashed, dotted, and dash-dotted grey lines indicate 1, 3, and 5 times the central thermal noise of the map ($\sigma = 1.6\,\mu$Jy\rev{/beam}), respectively. \rev{Our deblended counts show excellent agreement with the independent P(D) analysis down to the $3\sigma$ limit.} The solid vertical red line indicates the upper flux limit of 50~$\mu$Jy; sources brighter than this were masked prior to deblending.}
\label{fig:source_counts_real}
\end{figure*}

The primary scientific product of this work is the 1.4 GHz radio source number counts. The flux densities have been converted from the native MIGHTEE observing frequency to 1.4\,GHz using the spatially varying frequency-response maps of \cite{Hale2025}, adopting a spectral index $\alpha=0.7$. In Figure \ref{fig:source_counts_real}, we present the Euclidean-normalized source counts derived from our \textit{Comprehensive} and \texttt{Radio-Likely Prior} catalogues. Our primary result, from the \texttt{Radio-Likely Prior}, shows excellent agreement with the literature at certain flux ranges. At the bright end of our deblended range, from approximately 20 to 40 $\mu$Jy, our counts lie below the completeness-corrected literature values from \cite{Hale2025}. This turnover is expected from the 50~$\mu$Jy mask and a similar behavior is seen in the simulations (see Figure \ref{fig:source_counts_sim}), confirming that this is a feature of our methodology, not a discrepancy in the measurement.

Moving deeper into the confusion-dominated regime in MIGHTEE, from approximately 15 $\mu$Jy down to our faintest reliable bin at $\sim$4.8 $\mu$Jy (3$\sigma$), our counts show excellent agreement with the P(D) analysis of the MeerKAT DEEP2 field from \cite{Matthews2021}.
The agreement between our catalog-based deblended counts and this independent statistical measurement validates our methodology in a flux density regime where traditional source-finding fails due to confusion. This implies that we are resolving the vast majority of the radio background at these flux densities.

Below this 3$\sigma$ limit, our counts decline and fall below the P(D) result. This decline is expected since, based on the analysis of the simulations, below 3$\sigma$ the flux densities of the \texttt{Radio-Likely Prior} source counts systematically and increasingly underestimate the ground truth. Although we note that the only observational constraints on the true form of the source counts at these flux densities come from the P(D) analysis of \cite{Matthews2021}, where optical counterparts are not required.

In contrast, the counts derived from the sub-optimal \textit{Comprehensive L\textsubscript{IR} Prior} (shown in  Figure~\ref{fig:source_counts_real}, see Appendix \ref{appendix:lir} )
are systematically higher below 5$\sigma$. This result from real data confirms the prediction from our simulations: an inclusive but low-purity prior leads to an over-deblending effect that artificially boosts the number of faint sources. Intriguingly, at the faintest flux limits ($\sim$1--2$\sigma$), these overestimated counts from the \textit{Comprehensive Prior} appear to align with the P(D) results. This provides real-data confirmation of our primary simulation finding: an unfiltered prior will indeed generate a large population of spurious sources, but in doing so, it can statistically approximate the true underlying source population. This highlights both the remarkable accuracy of P(D) analysis in the confusion-dominated regime and the complementary power of our deblending method, which provides reliable flux densities for a pure, albeit incomplete, sample of individual sources.

\section{Conclusions and Future Work}
\label{sec:conclusions}

Deep radio continuum surveys from highly sensitive facilities like MeerKAT provide unprecedented access to the faint radio source population, enabling important tests of galaxy evolution models. However, their scientific potential is fundamentally limited by source confusion in the absence of appropriate analysis techniques. We have developed, validated, and applied a complete framework for producing high-fidelity deblended radio source catalogues from confusion-limited survey data, using the MIGHTEE-COSMOS field as our primary application. Our key findings and contributions are:

\begin{enumerate}
    \item {Deblending is essential for deep confused radio surveys:} We confirm through simulations that standard source-finding suffers from significant flux boosting in the confused regime, making it unsuitable for accurate faint-source science.

     \item {XID+ is an effective tool for radio deblending:} We have adapted and validated the probabilistic Bayesian framework XID+ for the specific challenges of deep radio continuum data. When configured with an appropriate prior list and masking strategy, it accurately recovers both individual source flux densities and the statistical distribution of the underlying source population.
    
    \item {Prior quality governs deblending accuracy:} The reliability of deblended flux densities is sensitive to the purity of the prior list. Our simulations and real-data application both show that a high-purity, albeit incomplete, prior (our \texttt{Radio-Likely Prior} ) yields far more precise and accurate flux measurements (with a scatter of $\approx$ 0.3 dex at 30 $\mu$Jy in simulations). In contrast, a complete but unfiltered prior leads to over-deblending, resulting in large, unreliable flux density uncertainties (>1 dex scatter) and a proliferation of spurious faint detections.   
    
    \item This work delivers the deepest catalog-based MIGHTEE-COSMOS source counts to date. When applied to the MIGHTEE-COSMOS field, our validated strategy recovers the 1.4 GHz source number counts down to the $3\sigma$ limit ($4.8\,\mu$Jy). The final catalogue contains 89,562 sources, including a recommended high-fidelity subset of 20,757 sources.
    
    \item Bayesian diagnostics and sampler optimization are important for ensuring catalogue fidelity: The Bayesian p-value residual statistic is a powerful tool for identifying sources with potentially unreliable flux measurements. We find that its successful application to complex, real-world data requires careful tuning of the MCMC sampler settings within the Stan framework. Increasing the \texttt{max\_treedepth} from the default of 10 to 15 was required for convergence on robust solutions, though with increased computational cost. This demonstrates that the increased complexity of real data requires a more exhaustive exploration of the posterior parameter space to achieve reliable fits.
\end{enumerate}

This deblending framework enables a wide range of science applications beyond what is possible with traditional source-finding. These include constructing a complete, dust-unbiased census of the {cosmic star formation history} (e.g.,\citealt{Novak2017, Matthews2024}); probing the evolution of the {infrared--radio correlation} to unprecedented depths (e.g.,\citealt{Delhaize2017, Delvecchio2021, Moloko2025}); disentangling star formation and AGN activity in {radio-quiet AGN} (e.g.,\citealt{Padovani2016, Panessa2019}) and precisely measuring the faint-end slope of the {radio luminosity function} (e.g.,\citealt{Smolcic2017, Algera2020}).
The framework established in this work is directly applicable to other confusion-limited radio surveys from current facilities (ASKAP, LOFAR, VLA, uGMRT). 

While the framework established in this work provides a validated approach for producing deep, reliable deblended radio source catalogues, it could be refined further. 

Specifically, in the current framework, we assume that the sources that lie at these faint flux densities are both unresolved (a reasonable assumption) and that their flux density is distributed over the restoring beam. However, the maps that we use are not cleaned to the levels that we are extracting the flux densities from, as such, we should use the synthesised beam as our PSF rather than the restoring beam, whose information is in the headers. However, this would introduce additional complexity without a significant amount of gain. The synthesised beam for the MIGHTEE data is tapered to produce a Gaussian-like profile that is very similar to the adopted restoring beam. As such, the difference between using the restoring beam approximation and the full synthesised beam is of the order 5~per cent, significantly less than the uncertainties on individual sources. However, ideally, one would simulate the full observational pipeline: Using the improved sky model, one could simulate the entire data acquisition and reduction process. This involves generating the raw instrumental visibilities, introducing realistic thermal noise and calibration errors, and then processing these data through the identical imaging and deconvolution pipeline used for the real MIGHTEE observations. This approach would naturally produce complex artefacts, such as residual sidelobes and spatially correlated noise, providing the ultimate test of this methodology against real-world instrumental effects. However, this would be very computationally intensive.

One could also improve the \texttt{Radio-Likely Prior} with more advanced source selection. The success of our deblending is highly dependent on the purity of the \texttt{Radio-Likely Prior}. While effective, the current method based on the IRRC can be advanced by using more sophisticated predictive models. Future work will explore the use of full-SED models to measure the SFR and/or machine learning algorithms trained on the rich, multi-wavelength photometry available in the COSMOS field, similar to \citep{Wang2024b} to learn the complex relationship between a galaxy's full UV-to-mid-infrared spectral energy distribution and its expected 1.4\,GHz flux. This would allow us to generate a more nuanced, probabilistic radio flux estimate for every source in the parent catalogue, creating a more robust and complete \texttt{Radio-Likely Prior} that better distinguishes detectable radio sources from undetectable ones.

\section*{Acknowledgements}
We thank the referee for useful suggestions which improved the
quality of the paper.
EM acknowledges financial support from the University of South Africa (Research Fund: 409000). MJJ, CLH and IHW acknowledges support from the Hintze Family Charitable Foundation through the Oxford Hintze Centre for Astrophysical Surveys. MJJ acknowledges the support of the STFC consolidated grant [ST/S000488/1] and [ST/W000903/1] and from a UKRI Frontiers Research Grant [EP/X026639/1]. MGS acknowledges support from the South African National Research Foundation (Grant No. 84156). DJBS acknowledges support from the UK's Science and Technology Facilities Council via grant ST/Y001028/1, and from the Leverhulme Trust via Research Project Grant RPG-2025-078. CLH acknowledges support
from STFC through grant ST/Y000951/1. JA acknowledges financial support from the Science and Technology Foundation (FCT, Portugal) through research grants UIDB/04434/2020 (DOI: 10.54499/UIDB/04434/2020), UIDP/04434/2020 (DOI: 10.54499/UIDP/04434/2020) and UID/04434/2025. MV acknowledges financial support from the Inter-University Institute for Data Intensive Astronomy (IDIA), a partnership of the University of Cape Town, the University of Pretoria and the University of the Western Cape, and from the South African Department of Science and Innovation's National Research Foundation under the ISARP RADIOMAP Joint Research Scheme (DSI-NRF Grant Number 150551) and the CPRR HIPPO Project (DSI-NRF Grant Number SRUG22031677). SJ acknowledges the European Union’s Horizon Europe research and innovation program under the Marie Sk\l{}odowska-Curie Action grant No. 101060888, and the Villum Fonden research grants 37440 and 13160. FXA acknowledges the support from the National Natural Science Foundation of China (12303016) and the Natural Science Foundation of Jiangsu Province (BK20242115). The authors thank Suman Chatterjee, Antonio la Marca, and Lingyu Wang for helpful discussions that enhanced this work. 

 The MeerKAT telescope is operated by the South African Radio Astronomy Observatory, which is a facility of the National Research Foundation, an agency of the Department of Science and Innovation. We acknowledge the use of the ilifu cloud computing facility – www.ilifu.ac.za, a partnership between the University of Cape Town, the University of the Western Cape, Stellenbosch University, Sol Plaatje University and the Cape Peninsula University of Technology. The ilifu facility is supported by contributions from the Inter-University Institute for Data Intensive Astronomy (IDIA – a partnership between the University of Cape Town, the University of Pretoria and the University of the Western Cape), the Computational Biology division at UCT and the Data Intensive Research Initiative of South Africa (DIRISA). The authors acknowledge the Centre for High Performance Computing (CHPC), South Africa, for providing computational resources to this research project. This work made use of the CARTA (Cube Analysis and Rendering Tool for Astronomy) software (DOI 10.5281/zenodo.3377984 – https://cartavis.github.io). EM acknowledges the use of AI-based tools for assistance with code development and for improving the clarity and language of the manuscript.

\section*{Data Availability}
The MIGHTEE DR1 data products are available at \hyperlink{SARAO archive}{https://doi.org/10.48479/7msw-r692}. The deblended radio source catalogues (the Radio-Likely and Comprehensive samples) generated in this work are available in a VizieR catalogue available at CDS via anonymous ftp to cdsarc.u-strasbg.fr (130.79.128.5) or via  \hyperlink{VizieR}{https://cdsarc.cds.unistra.fr/viz-bin/cat/J/MNRAS}. 
\bibliographystyle{mnras} 
\bibliography{mnras.bib} 

\appendix

\section{Corner Plots for Star-Galaxy Classification}
\label{app:corner_plots}

\begin{figure*}
\centering
\includegraphics[width=\textwidth]{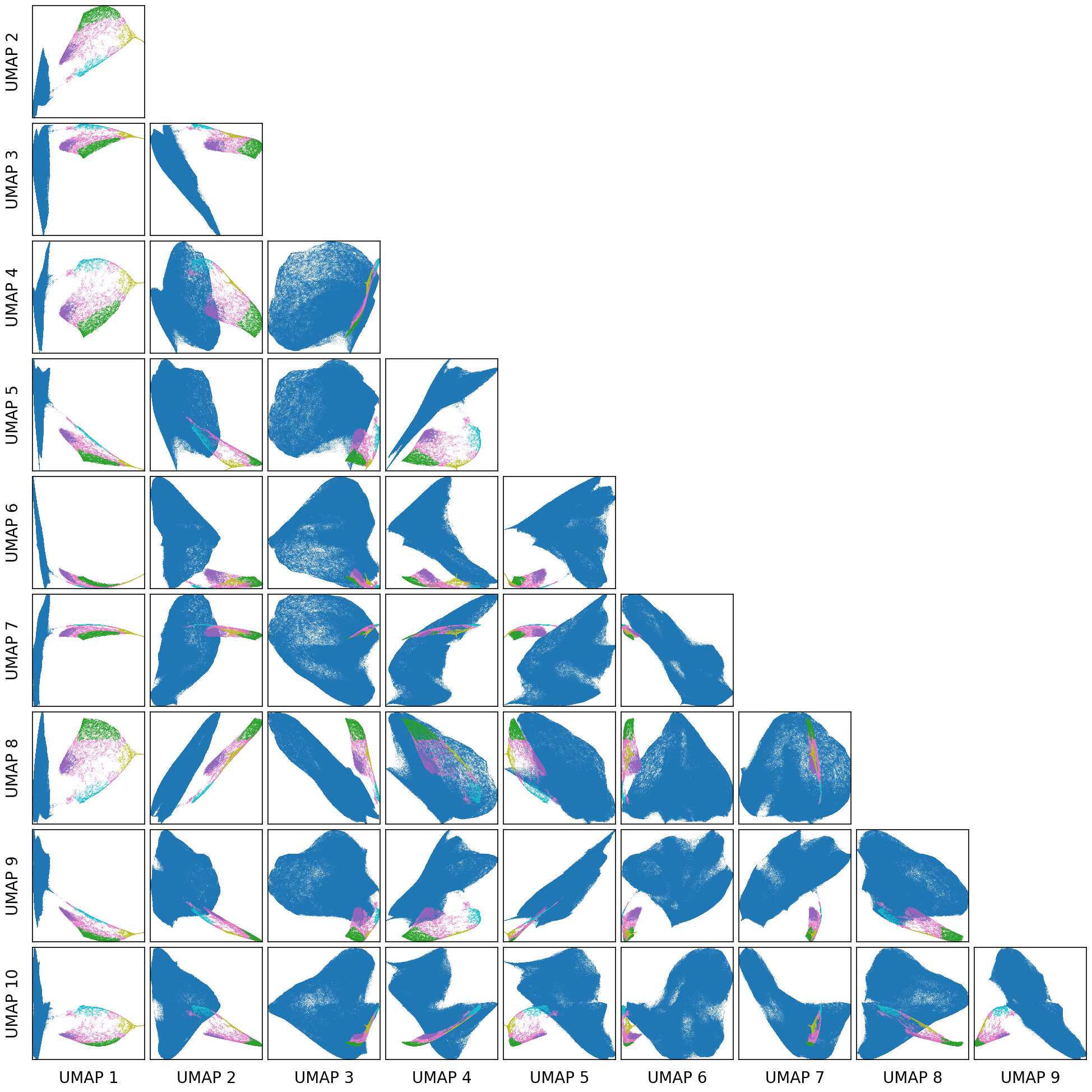}
\caption{Corner plot of the 10 UMAP dimensions for the HSC+UltraVista dataset. The colours are consistent with Figure \ref{fig:umap_hsc}, showing how clusters separate across different dimensional pairings.}
\label{fig:corner_hsc_appendix}
\end{figure*}

To supplement the two-dimensional UMAP projections shown in the main text, Figure \ref{fig:corner_hsc_appendix} shows the corner plot for all the 10-dimensional reduced feature space for the  HSC+VISTA dataset. This plot illustrate how the separation between the stellar and galactic clusters is maintained across the various pairings of the reduced UMAP dimensions, reinforcing the robustness of the classification.

\section{Deblending Performance with GAUL-based Masking}
\label{app:gaul_sims}

To supplement the primary analysis presented in Section \ref{sec:srl_vs_gaul}, we also investigated the impact of using a bright-source mask generated from the PyBDSF \texttt{GAUL} catalogue instead of the \texttt{SRL} (source list) catalogue. The \texttt{GAUL} catalogue treats every detected Gaussian component individually, making it a useful alternative to test the robustness of our masking strategy.

\begin{figure*}
    \centering
    \includegraphics[width=0.98\textwidth]{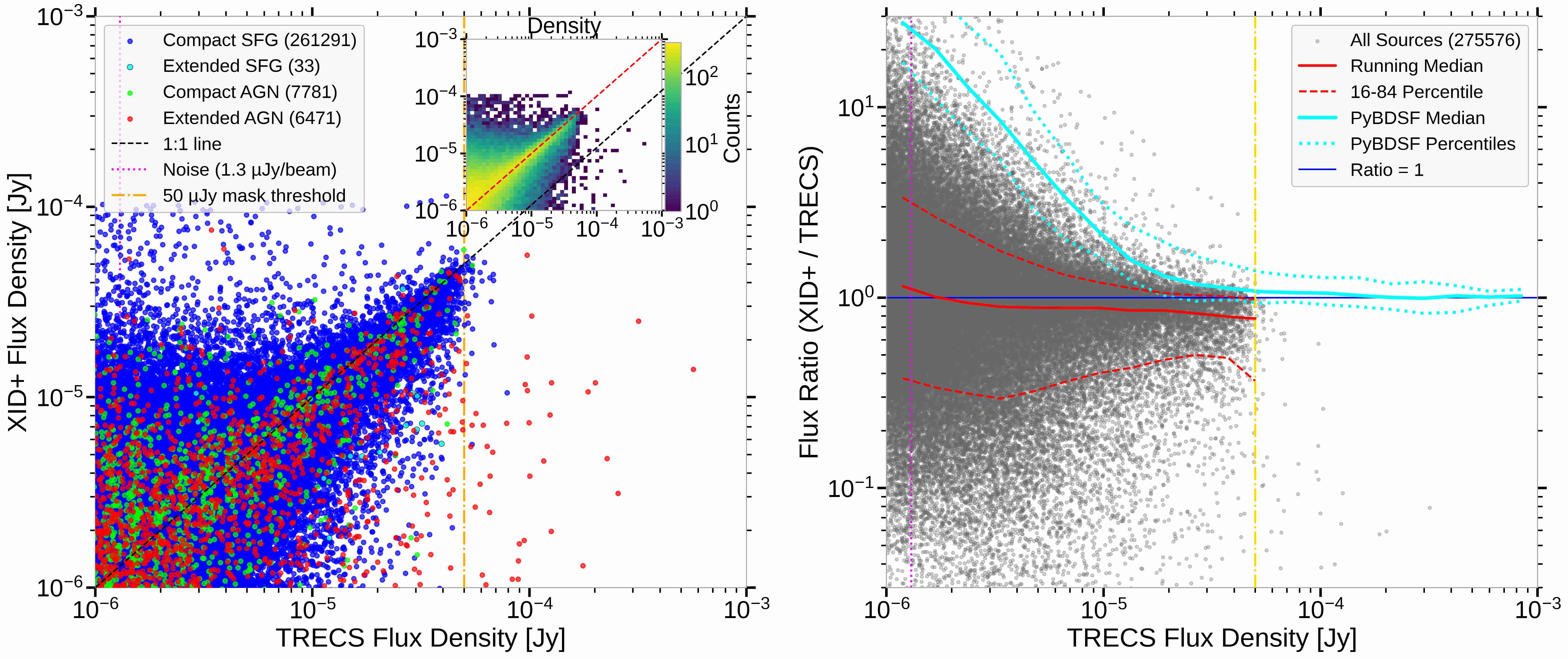}\\
    \includegraphics[width=0.98\textwidth]{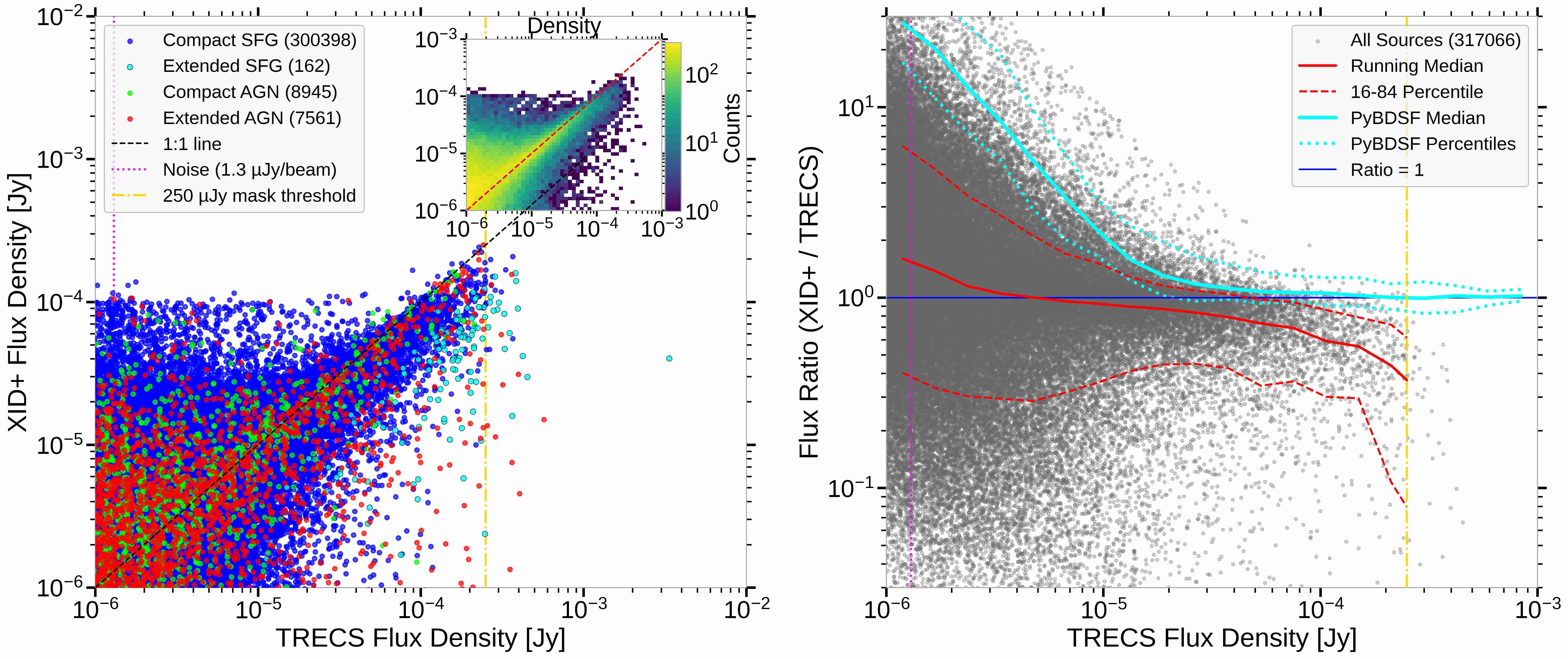}
    \caption{XID+ deblending performance using the \textbf{full T-RECS prior} with masks derived from the PyBDSF \textbf{GAUL} catalogue. The panels and symbols are the same as in Figure \ref{fig:sim_flux_comp_full}. Left: Performance when masking sources brighter than 50~$\mu$Jy using the GAUL catalogue. Right: Performance when masking sources brighter than 250~$\mu$Jy using the GAUL catalogue. The results are qualitatively identical to the SRL-masked case shown in Figure \ref{fig:sim_flux_comp_full}, with the 50~$\mu$Jy mask yielding lower scatter.}
    \label{fig:sim_flux_comp_full_gaul}
\end{figure*}

\begin{figure*}
    \centering
    \includegraphics[width=0.98\textwidth]{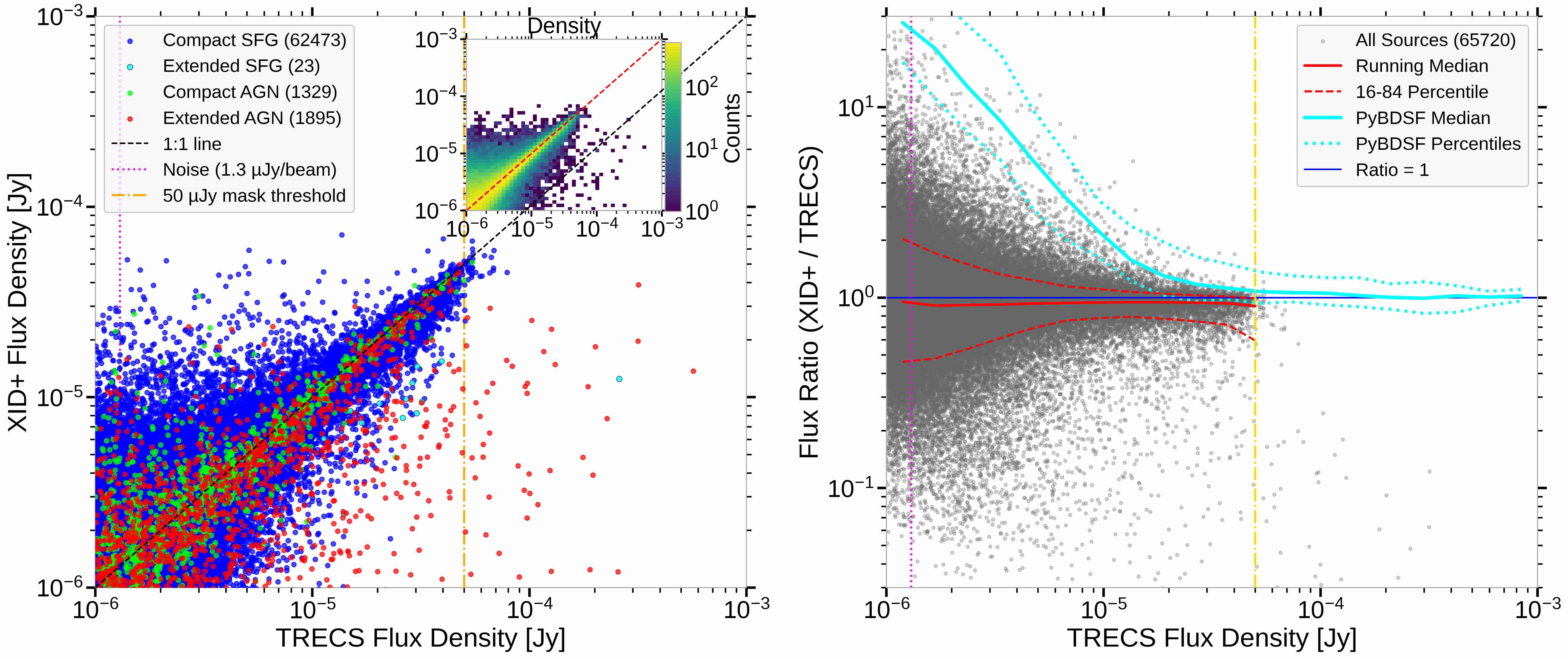}\\
    \includegraphics[width=0.98\textwidth]{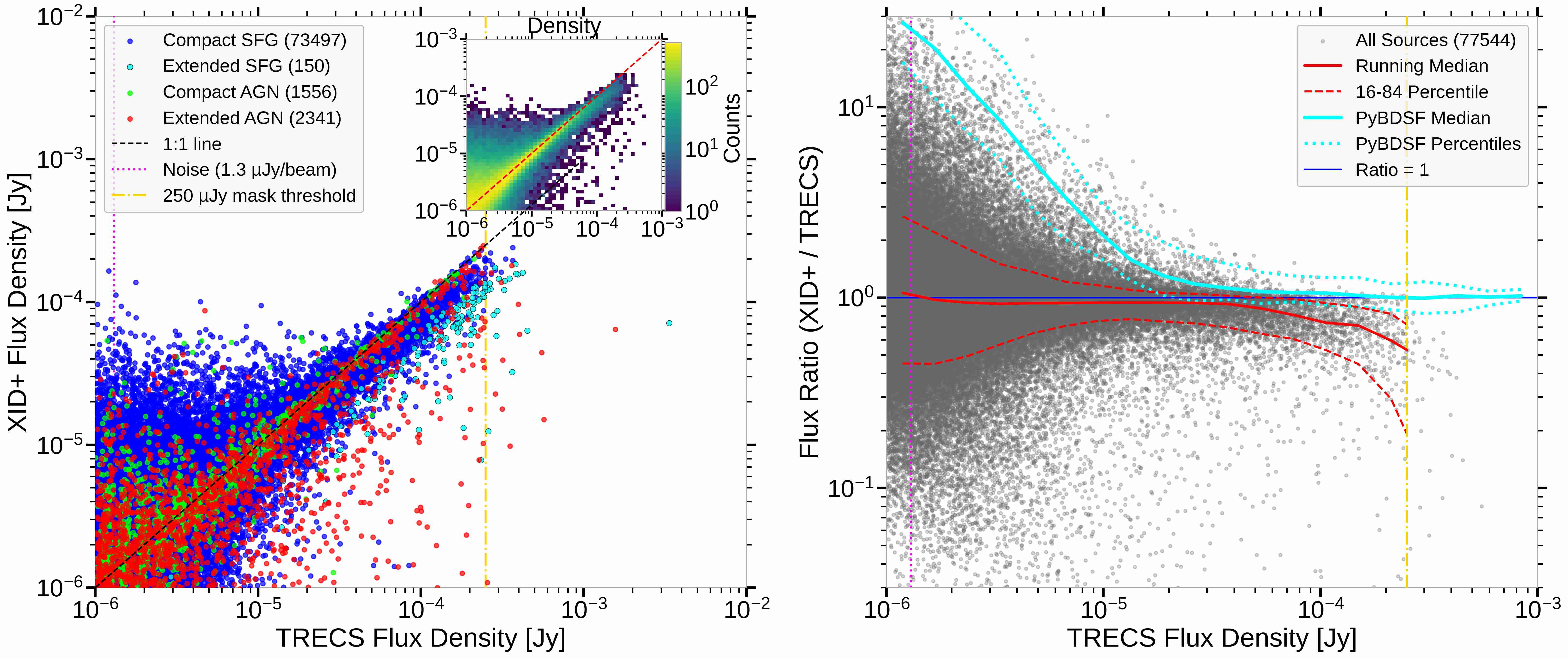}
    \caption{XID+ deblending performance using the realistic, incomplete \texttt{Radio-Likely Prior}  with masks derived from the PyBDSF \textbf{GAUL} catalogue. The panels and symbols are the same as in Figure \ref{fig:sim_flux_comp_full}. Left: Performance with a 50~$\mu$Jy GAUL mask. \textit{Right Plot:} Performance with a  250~$\mu$Jy GAUL mask. As with the SRL-masked case (Figure \ref{fig:sim_flux_comp_fir}), the 50~$\mu$Jy mask produces the most reliable flux measurements with the tightest correlation.}
    \label{fig:sim_flux_comp_fir_gaul}
\end{figure*}

The deblending performance plots for this test are presented in Figures \ref{fig:sim_flux_comp_full_gaul} and \ref{fig:sim_flux_comp_fir_gaul}. The results are qualitatively and quantitatively very similar to those derived using the \texttt{SRL} mask in the main text. This confirms that our main conclusion, that the quality of the prior list and the $>$ 50~$\mu$Jy masking threshold are the dominant factors driving the fidelity of the deblended flux densities is robust and not sensitive to the specific catalogue type used to generate the bright-source mask.

\section{Deblending with a \textit{Comprehensive L\textsubscript{IR} Prior}}
\label{appendix:lir}

To provide a real-world validation of our simulation results (Sec. \ref{sec:sim_validation}), we also performed a deblending run on the MIGHTEE data using \textit{Comprehensive L\textsubscript{IR} Prior}. This prior was designed to be as inclusive as possible, containing 298,894 potential sources after masking, and is analogous to the \texttt{Full Prior} used in our simulations. Using the same optimized sampler settings as for the \texttt{Radio-Likely Prior} (\texttt{max\_treedepth} up to 15), we find that 6,675 sources have a poor fit (p-value > 0.6), a dramatic reduction from the 22,133 flagged in the initial run. However, 
even with optimized settings, a prominent cloud of sources with significantly overestimated flux densities remains when compared to the independent super-deblended catalogue. While the fraction of flagged sources has dropped to $\sim$2.2 per cent of the total sample, the overall scatter in the flux comparison is visibly larger than in the \texttt{Radio-Likely Prior} case.

The most important outcome is seen in the source counts (Figure \ref{fig:source_counts_real}). The source counts (Figure \ref{fig:source_counts_real}) derived from this comprehensive prior are systematically higher than both our final \texttt{Radio-Likely} counts and the literature values. Likely due to over-deblending, where the algorithm incorrectly assigns flux to numerous faint, undetectable priors, artificially boosting the number of faint sources. 

This exercise demonstrates that the choice of prior is not merely a minor technical detail but a critical factor that directly impacts the final scientific conclusions. It validates our decision to focus the main body of this paper on the results from the carefully constructed, high-purity \texttt{Radio-Likely Prior} .

\begin{figure}
\centering
\includegraphics[width=\columnwidth]{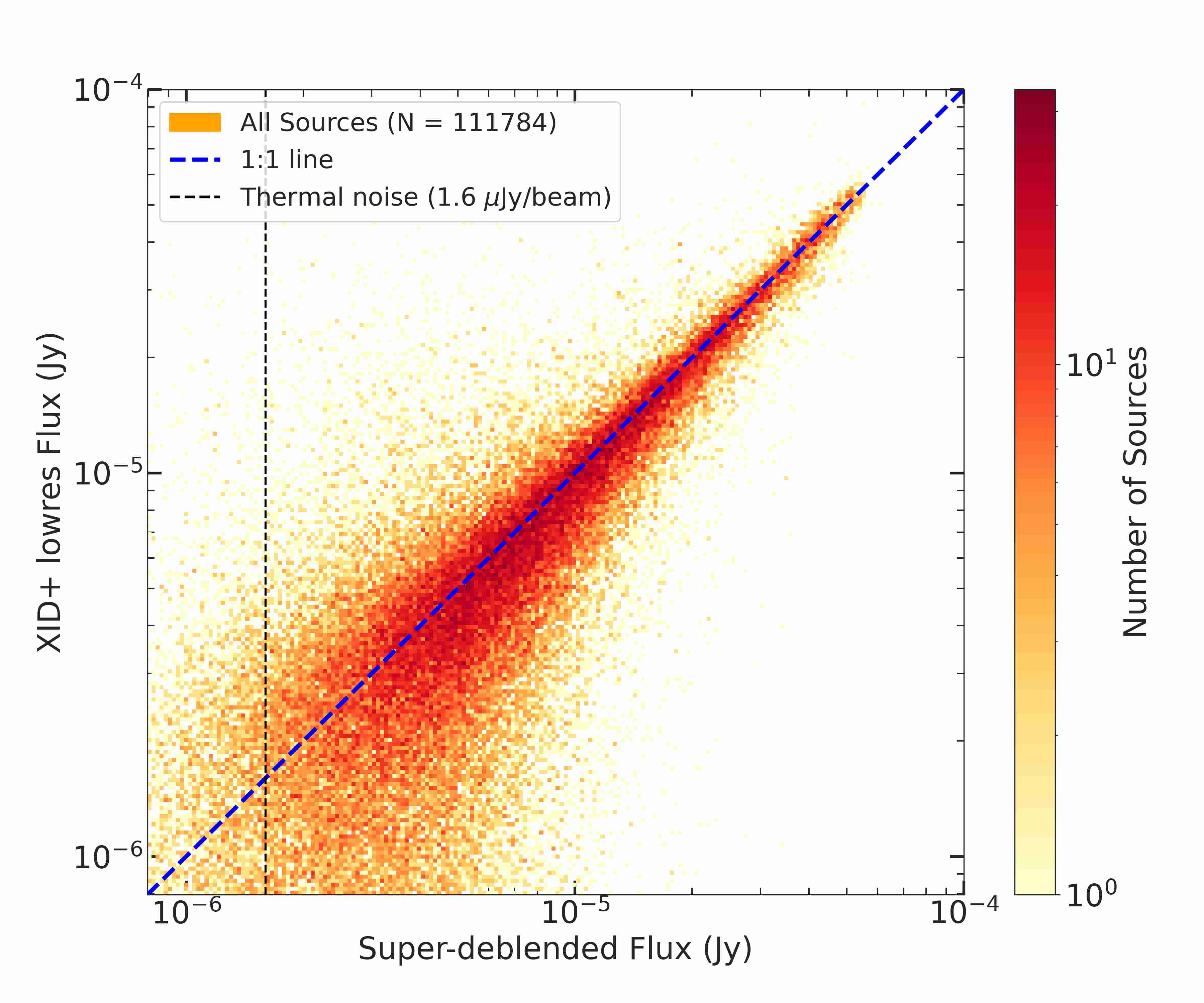}
\caption{Comparison of XID+ deblended flux densities against the super-deblended flux densities for the unfiltered sample derived using the \textit{Comprehensive L\textsubscript{IR} Prior}.}
\label{fig:flux_comp_full_LIR_appendix}
\end{figure}

\begin{figure}
\centering
\includegraphics[width=\columnwidth]{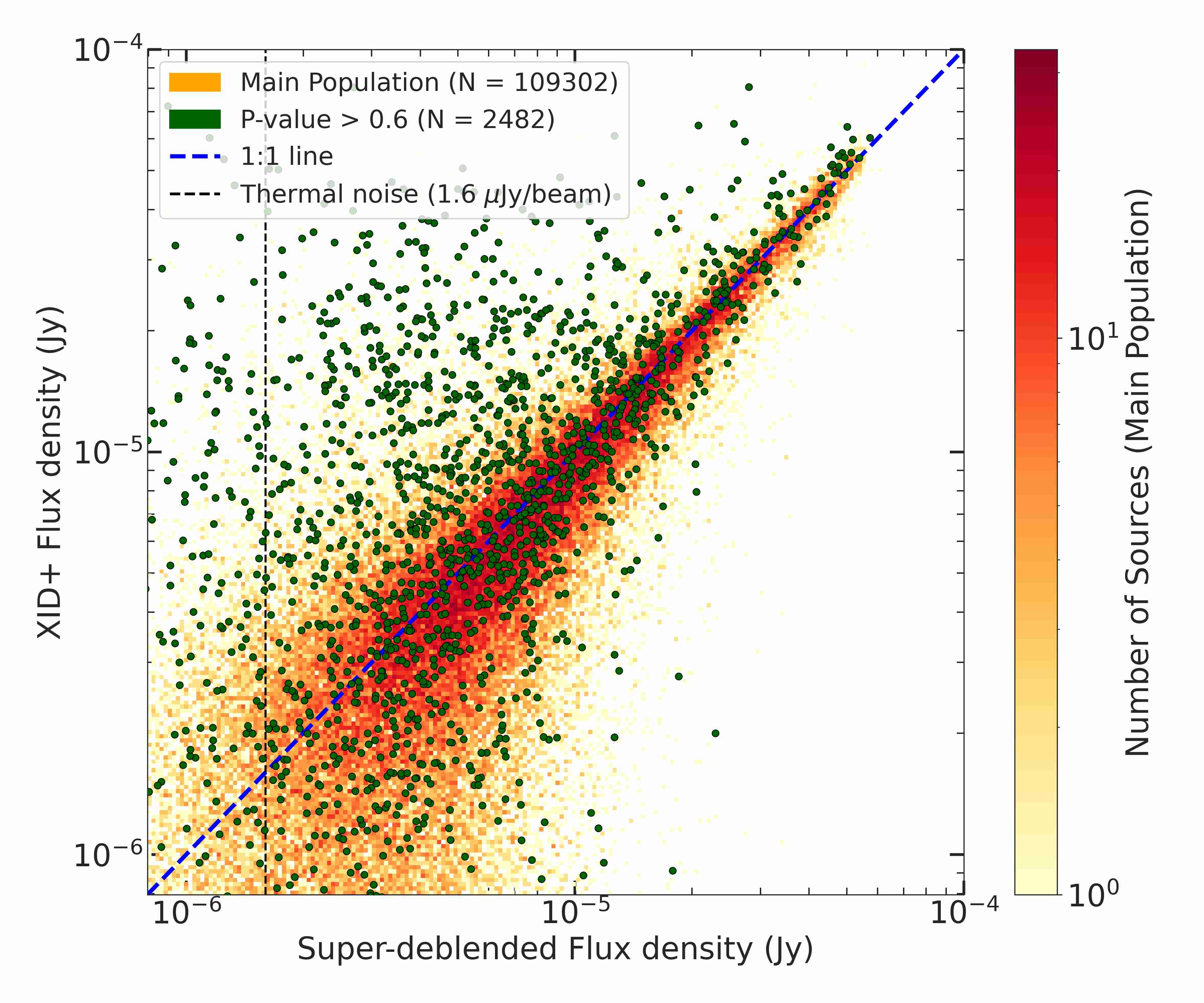}
\caption{The same as Figure \ref{fig:flux_comp_full_LIR_appendix}, but with sources having an unreliable fit ($p$-value > 0.6) highlighted.}
\label{fig:pvalue_highlighted_LIR}
\end{figure}

\begin{figure}
\centering
\includegraphics[width=0.8\columnwidth]{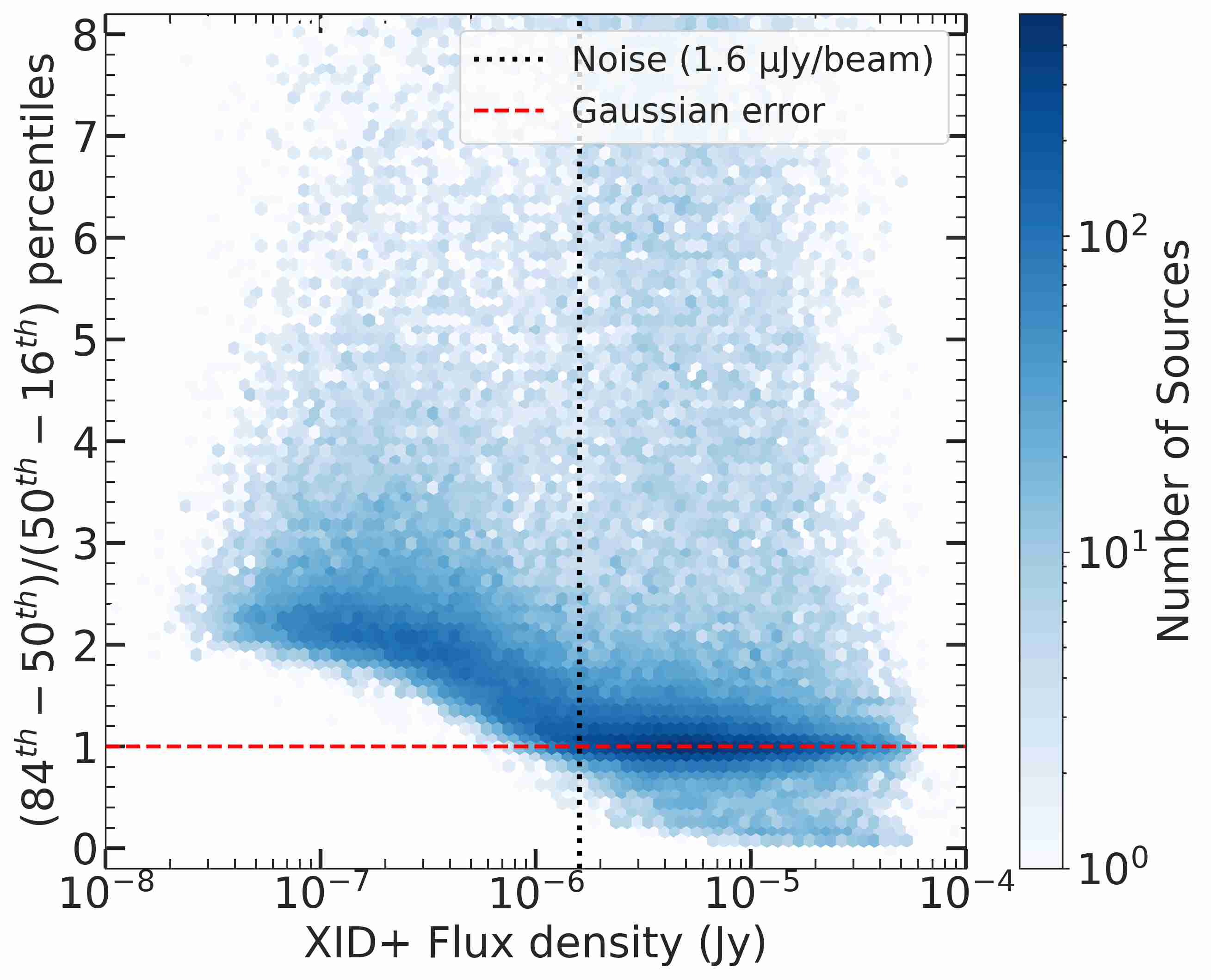}
\caption{The ratio of the upper to lower flux density uncertainty as a function of the deblended XID+ flux for sources from the \textit{Comprehensive L\textsubscript{IR} Prior}.}
\label{fig:uncertainty_LIR}
\end{figure}

\section{Supplementary Data Tables}
\label{app:data_tables}

The data tables for the derived source counts and a description of the final deblended source catalogue.

\subsection{Radio Source Counts Data}

The Euclidean-normalized source counts derived from the final, quality-filtered \texttt{Radio-Likely Prior} catalogue are provided in Table \ref{tab:counts_radio_likely}. These data points are plotted in Figure \ref{fig:source_counts_real}.

\begin{table}
    \centering
    \caption{Euclidean-normalized source counts for the MIGHTEE XID+ Radio-Likely prior catalogue.}
    \label{tab:counts_radio_likely}
    \begin{tabular}{cccc}
        \hline\hline
        Flux Bin & Number of & Source Counts & 68 per~cent CI Region \\
        (mJy) & Sources & (Jy$^{1.5}$ sr$^{-1}$) & (Jy$^{1.5}$ sr$^{-1}$) \\
        \hline
        1.000 -- 1.585 & 5900 & $0.04 $ & $[0.02, 0.04]$ \\
        1.585 -- 2.512 & 8166 & $0.12 $ & $[0.07, 0.11]$ \\
        2.512 -- 3.981 & 11379 & $0.33$ & $[0.19, 0.31]$ \\
        3.981 -- 6.309 & 13548 & $0.78$ & $[0.50, 0.74]$ \\
        6.309 -- 10.000 & 11772 & $1.35$ & $[1.08, 1.45]$ \\
        10.000 -- 16.000 & 8627 & $1.9$  & $[1.59, 3.02]$ \\
        16.000 -- 25.000 & 4899 & $2.3$   & $[1.98, 6.18]$ \\
        25.000 -- 40.000 & 2484 & $2.22$  & $[1.97, 7.99]$ \\
        \hline
    \end{tabular}
\end{table}

\subsection{Deblended Catalogue Format}

The format and column descriptions for the final deblended MIGHTEE radio point source catalogue are provided in Table \ref{tab:catalog_columns}.

\begin{table*}
	\centering
	\caption{Columns contained in our XID+ deblended MIGHTEE source catalogue.}
	\label{tab:catalog_columns}
	\begin{tabular}{lll}
		\toprule
		\textbf{Name} & \textbf{Unit} & \textbf{Description} \\
		\midrule
		\texttt{ID} & - & The COSMOS2020 ID \\
		\texttt{R.A.} & - & Right Ascension from COSMOS2020 \\
		\texttt{Dec.} & - & Declination from COSMOS2020 \\
		\midrule
		\texttt{Flux\_med} & mJy & MIGHTEE 1.3 GHz flux density (median) \\
		\texttt{Flux\_MAP} & mJy & MIGHTEE 1.3 GHz flux density (Marginal MAP from KDE) \\
		\texttt{Flux\_Err\_u} & mJy & MIGHTEE 1.3 GHz flux density (84th Percentile) \\
		\texttt{Flux\_Err\_l} & mJy & MIGHTEE 1.3 GHz flux density (16th Percentile) \\
		\texttt{bkg} & mJy/Beam & Fitted Background of MIGHTEE 1.3 GHz map (median) \\
		\texttt{Rhat} & - & Convergence Statistic (ideally < 1.2) \\
		\texttt{n\_eff} & - & Number of effective samples (ideally > 40) \\
		\texttt{P-value\_res} & - & Bayesian p-value residual statistic (ideally < 0.6) \\
        \texttt{Significance} & - & Detection significance (ideally $\> 3$) \\
		\bottomrule
	\end{tabular}
\end{table*}

\end{document}